\documentclass[12pt]{article}
\usepackage{amsmath}
\usepackage{graphicx}
\usepackage{enumerate}
\usepackage{natbib}
\usepackage{algorithm}
\usepackage{algorithmic}
\usepackage{graphicx}
\usepackage{epstopdf}
\usepackage{bm}
\usepackage{color,xcolor}
\usepackage{float}
\usepackage{url} 
\usepackage{subfig}
\usepackage[colorlinks,citecolor=blue,urlcolor=blue]{hyperref}

\usepackage{amsmath,verbatim,amssymb}
\usepackage{verbatim,amsmath}
\usepackage{lscape}
\usepackage[toc,page,title,titletoc,header]{appendix}

\newtheorem{theorem}{Theorem}
\newtheorem{proposition}{Proposition}
\newtheorem{corollary}{Corollary}

\renewcommand{\hat}{\widehat}

\newcommand{\bse}{\begin{eqnarray*}}
\newcommand{\ese}{\end{eqnarray*}}

\def\boxit#1{\vbox{\hrule\hbox{\vrule\kern6pt
          \vbox{\kern6pt#1\kern6pt}\kern6pt\vrule}\hrule}}

\usepackage{tikz}
\usetikzlibrary{shapes.geometric, arrows}
\usetikzlibrary{calc}
\usetikzlibrary{positioning, shapes.geometric}
\usetikzlibrary{arrows,shapes,chains}

\usepackage{amsfonts}
\usepackage[letterpaper,  margin=1in]{geometry}
\renewcommand{\baselinestretch}{1.5}
\newcommand{\blind}{0}

\begin{document}

\def\spacingset#1{\renewcommand{\baselinestretch}%
{#1}\small\normalsize} \spacingset{1}


\if0\blind
{
  \bigskip
  \bigskip
  \bigskip
  \begin{center}
    {\LARGE\bf Inference on High-dimensional Single-index Models
with Streaming Data}

\bigskip

Dongxiao Han$^1$, Jinhan Xie$^2$, Jin Liu$^1$, Liuquan Sun$^3$, Jian Huang$^{4}$, Bei Jiang$^2$ and Linglong Kong$^2$

\bigskip

\footnotetext[2]{Dongxiao Han and Jinhan Xie are co-first authors.
}


{
$^1$School of Statistics and Data Science, LPMC, KLMDASR and LEBPS, Nankai University, Tianjin, China

$^2$Department of Mathematical and Statistical Sciences, University of Alberta, Edmonton,
Canada

$^3$Institute of Applied Mathematics, Academy of Mathematics and Systems Science, Chinese Academy of Sciences, Beijing, China

 $^4$Department of Applied Mathematics, The Hong Kong Polytechnic University,
 Hong Kong, China



}

\end{center}
  \medskip
} \fi

\bigskip
\begin{abstract}
{ 
Traditional statistical methods are faced with new challenges due to streaming data. The major challenge is the rapidly growing volume and velocity of data, which makes storing such huge datasets in memory impossible. The paper presents an online inference framework for regression parameters in high-dimensional semiparametric single-index models with unknown link functions. The proposed online procedure updates only the current data batch and summary statistics of historical data instead of re-accessing the entire raw data set. At the same time, we do not need to estimate the unknown link function, which is a highly challenging task. In addition, a generalized convex loss function is used in the proposed inference procedure. To illustrate the proposed method, we use the Huber loss function and the logistic regression model's negative log-likelihood. In this study, the asymptotic normality of the proposed online debiased Lasso estimators and the bounds of the proposed online Lasso estimators are investigated. To evaluate the performance of the proposed method, extensive simulation studies have been conducted. We provide applications to Nasdaq stock prices and financial distress datasets.
}

\end{abstract}

\noindent%
{\it Keywords:} {High-dimensional data; Lasso; Single-index models; Statistical inference; Streaming data.}
\vfill

\newpage
\spacingset{1.5} 
\section{Introduction}
\label{sec:intro}

The rapid development of data collection techniques brings new challenges to develop online approaches to the data in a streaming fashion. In such a data environment, it is often numerically challenging or sometimes infeasible to store the entire dataset in memory. Consequently, the classical offline methods which involve the entire dataset are less attractive or  even infeasible due to computationally expensive. Instead, online methods can be used to process the out-of-memory data and make real-time decisions, which have been prevalent in economics, finance, machine learning, and statistics. Up to now, various online methods have been proposed. For example, the stochastic gradient descent (SGD) algorithm and its variants have been extended to the streaming settings; see \citet{duchi2009efficient}, \citet{xiao2009dual}, \citet{dekel2012optimal}, \citet{chen2020statistical}, \citet{chen2021online} and \citet{zhu2021online}. In addition,
\citet{lin2011aggregated} considered an aggregated estimating equation for generalized linear models. \citet{schifano2016online} proposed  online-updating algorithms and inferences applicable for linear models and
estimating equations. \citet {luo2020renewable} suggested  a renewable estimation and incremental inference to analyze streaming data sets using generalized linear models. The aforementioned online methods are developed for
low-dimensional settings where the number of regressors is fixed and far less than the total sample size.

In recent years, a large amount of high-dimensional
data streams, such as network flows, wireless sensor networks data, and multimedia streams have been generated; see \citet{wang2017robust}, \citet{braverman2017clustering}, and \citet{din2021learning}. To analyze the above high-dimensional data streams, many online methods have been studied. For example, \citet{langford2009sparse} proposed an online $\ell_1$-regularized method via a variant of
the truncated SGD. \citet{fan2018statistical} developed the diffusion approximation approach to investigate SGD estimators. \citet{gepperth2021gradient} presented an approach for Gaussian mixture model via SGD with non-stationary, high-dimensional streaming data. \citet{shi2021statistical} introduced a  valid inference method for single or low-dimensional regression coefficients via a recursive online-score estimation technique.
\citet{deshpande2021online} considered a class of online estimators in a high-dimensional
auto-regressive model. \citet{han2021online} proposed an online debiased lasso estimator for statistical inference with high-dimensional streaming data and further extended to the generalized linear models in \citet{luo2021statistical}. The above existing estimation and inference procedures only focused on the linear or generalized linear models. However,  much less is known under potential misspecification of these commonly used models or more general models.

The single-index models (SIMs), which accommodate possible nonlinearity and avoid the curse of dimensionality
simultaneously,  are useful extensions of the linear regression model. Over the last few decades, the SIMs
have been widely investigated in both the statistics and econometrics literature. In low-dimensional settings, the SIMs have been studied extensively in the literature, see \citet{carroll1997generalized}, \citet{xia2009adaptive}, and \citet{cui2011efm}, among others. In high-dimensional settings, the SIMs have also attracted interest with various studies such as variable selection, estimation, and hypothesis. For example, \citet{alquier2011sparse} introduced a PAC-Bayesian estimation approach for the sparse SIMs. \citet{ganti2015learning} provided a suite of algorithms to learn the SIMs.  \citet{radchenko2015high} proposed a non-parametric least squares with an equality $\ell_1$ constraint to simultaneous variable selection and estimation. Sign support
recovery for the regression coefficient vector was studied by \citet{neykov2016l1}.  \cite{yang2017high} considered the estimation problems of the parametric component of the SIMs.  \citet{zhang2020ultra} proposed flexible regularized  single-index quantile regression models for high-dimensional
data. \citet{eftekhari2021inference} conducted  pointwise inference based on least squares.  However, the existing estimation or inference methods of the SIMs have been studied on the fixed sample size before data collection and might not be suitable to implement the situation that data arrives in a streaming manner.

In this paper, we develop an online framework for real-time estimation and inference of regression parameters in SIMs with streaming data. Our proposed procedure is established based on general convex loss functions. We consider the Huber loss function \citep{huber1964robust} and the negative log-likelihood of logistic regression model as two special examples to illustrated the proposed method.  Unlike the previous works, the proposed online estimators are updated via the current data batch and summary statistics of historical data without accessing the entire raw dataset. Meanwhile, we do not need to estimate any unknown link functions at each stage. In addition, the proposed online method accounts for sparsity features in a candidate set of covariates and provides a valid statistical inference procedure for regression parameters. Under some regular conditions, we also show that the consistency and asymptotic normality of the proposed online estimators, which provides us a theoretical basis for carrying out real-time statistical inference with streaming data. \citet{han2021online} and \citet{luo2021statistical} also considered inference for high-dimensional models with streaming data. However, our work differs from theirs in the following two aspects: (i) The proposed procedure aims at the SIMs, while their methods focused on linear  and generalized linear models, respectively; (ii) Unlike the case of high-dimensional linear or generalized linear models where the loss function is assumed to be second-order differentiable, we only require the existence of the first-order derivative of the loss function. In particular, our general loss function includes the Huber loss as a special case, which is robust to response.

The rest of this paper is organized as follows. In Section 2.1, we present the model settings. The proposed
online estimation procedure with its theoretically property
is presented in Section 2.2.  Section 2.3 introduces the
proposed online one-step procedure. Some examples are provided to illustrate the proposed method in Section 3. We evaluate the performance of the proposed procedure through simulation studies in Section 4. In Section 5, we apply the proposed method to the Nasdaq stock and financial distress datasets. Some discussions are given in Section 6. Technical details are deferred to the supplementary material.

\section{Model and methodology}
\subsection{Single-index models}
We consider the following high-dimensional SIMs
\citep{neykov2016l1}:
\begin{align}
\label{E1}
Y=f(\boldsymbol{X}^\top\boldsymbol{\beta}_0,\epsilon), 
\end{align}
where $Y$ is a response variable, $\boldsymbol{X}$ is a $p$-dimensional covariate vector, $\boldsymbol{\beta}_0$ is a $p$-dimensional vector of regression parameters, $f$ is an unknown link function, and  $\epsilon$ is an error term
whose distribution is unspecified. Without loss of generality, we assume $E(\boldsymbol X)=0$.
Assume that $E(\boldsymbol\beta_0^\top\boldsymbol\Sigma \boldsymbol\beta_0)=1$
\citep{neykov2016l1,eftekhari2021inference} 
 for identifiability, where $\boldsymbol\Sigma=E(\boldsymbol X\boldsymbol X^\top)$. Consider a time point $m\ge 2$ with a total of $N_m=\sum_{j=1}^mn_j$ independent copies of $(Y,\boldsymbol X)$ arriving in a sequence of $m$ data batches, denoted by $\{\mathcal{D}_1,\dots,\mathcal{D}_m\}$, where $n_j$ is the size of the batch $\mathcal{D}_j$. For any $1\le j\le m$, denote the observations in $\mathcal{D}_j$ by
$\{Y_i^{(j)},\boldsymbol X_i^{(j)}\}_{i=1}^{n_j}$. The SIMs involve many existing models as special cases, such as the linear regression model and the logistic regression model.\\

\subsection{Online consistent estimation}

The recovery of $\boldsymbol\beta_0$ up to a scale under model (\ref{E1}) often depends on the linearity of expectation assumption 
\citep{li1989regression,li1991sliced,neykov2016l1} given below:

\noindent{\bf Definition 1  (Linearity of Expectation)} A $p$-dimensional random variable $\boldsymbol W$ is said to satisfy linearity of expectation in the direction of $\boldsymbol\beta$ if for any direction $\boldsymbol b\in\mathbb R^p$:
\begin{align*}
E(\boldsymbol W^\top \boldsymbol b|\boldsymbol W^\top\boldsymbol\beta)=c_{\boldsymbol b}\boldsymbol W^\top\boldsymbol\beta+a_{\boldsymbol b},
\end{align*}
where $a_{\boldsymbol b}$ and $c_{\boldsymbol b}$ are two constants which might depend on the direction $\boldsymbol b$.

Notice that commonly-used elliptically symmetric distributions \citep{fang1990statistical}
involving Gaussian distributions as
special cases satisfy the
linearity in expectation uniformly in all directions
\citep{cambanis1981theory}.
We consider estimating
$\boldsymbol\beta_0$ up to a scalar by using a loss
function $l(Y,\boldsymbol X^\top\boldsymbol\beta)$. The following condition is for the parameter identification.

\begin{itemize}
\item [(C1)] Assume that $\boldsymbol X$ is independent of $\epsilon$, and satisfies the linearity of expectation assumption in the direction of $\boldsymbol\beta_0$. In addition, suppose that the function $(Y,\boldsymbol X^\top\boldsymbol\beta)
    \rightarrow l(Y,\boldsymbol X^\top\boldsymbol\beta)$ is  convex in $\boldsymbol X^\top\boldsymbol\beta\in \mathbb{R}$, and the function
     $\boldsymbol\beta\rightarrow E\{l(Y,\boldsymbol X^\top\boldsymbol\beta)\}$  has a unique minimizer
     $\boldsymbol\beta^*\ne 0$.
\end{itemize}

Condition (C1) is a regular condition for the SIMs
\citep{li1989regression,li1991sliced,neykov2016l1}
The next Proposition \ref{P1} serves as the basis of our work.
\begin{proposition}
\label{P1}
Suppose that condition (C1) hold. Then there exists some non-zero constant $k_1$ depending on $l(Y,\boldsymbol X^\top\boldsymbol\beta)$  such that
$\boldsymbol\beta^*=k_1\boldsymbol\beta_0$.
\end{proposition}

Proposition \ref{P1} indicates that a consistent estimator of $\boldsymbol\beta_0$ up to a scalar can be obtained by
minimizing the following penalized empirical version of $E\{l(Y,\boldsymbol X^\top\boldsymbol\beta)\}$ under some mild condition:
\begin{align*}
\frac{1}{N_m}\sum_{j=1}^m\sum_{i=1}^{n_j}l(Y_i^{(j)},\boldsymbol X_i^{(j)\top}\boldsymbol\beta)+\lambda_n\|\boldsymbol\beta\|_1,
\end{align*}
where $\lambda_n$ is a tuning parameter, $\|\boldsymbol\beta\|_1=\sum_{l=1}^p|\beta_l|$ is the $\ell_1$-norm of $\boldsymbol\beta$, and
$\beta_l$ is the $l$th element of $\boldsymbol\beta$. However, under the streaming data setting, since new data arrives continually,
data volume accumulates very fast over time. This leads to the result that the
raw data can not be stored in memory for a long time and we can not access the entire dataset $\{\mathcal{D}_1,\dots,\mathcal{D}_m\}$ at the time point $m$,  making it impossible to implement the algorithm above.
To tackle this problem, we consider an online updating procedure which just exploit the current
data and the summary statistics from the historical raw data for estimating $\boldsymbol\beta^*$. To remove the dependence between an estimator of $\boldsymbol{\beta}^*$, and the
observed data, we employ a sample-splitting technique. Without loss of generality, assume that
$n_1,\dots, n_m$ are all even numbers.
Let $\mathcal{D}_{j,1}=\{Y_i^{(j)},\boldsymbol X_i^{(j)}\}_{i=1}^{n_j/2}$, and $\mathcal{D}_{j,2}=\{Y_i^{(j)},\boldsymbol X_i^{(j)}\}_{i=n_j/2+1}^{n_j}$, for $j=1,\dots,m$.
Define
\begin{align*}
\boldsymbol H=\frac{\partial^2 }{\partial{\boldsymbol\beta}
\partial{\boldsymbol\beta}^\top}E\{l(Y,\boldsymbol X^\top\boldsymbol\beta)\}|_{\boldsymbol\beta=\boldsymbol\beta^*}.
\end{align*}
When the batch $\mathcal{D}_1$ arrives, let $\hat{\boldsymbol{\beta}}_1^{(1)}$  be the minimizer of
\begin{align}
\label{E2}
&\frac{2}{n_1}\sum_{i=1}^{n_1/2}l(Y_i^{(1)},\boldsymbol X_i^{(1)\top}\boldsymbol\beta)+\lambda_{1}\|\boldsymbol\beta\|_1,
\end{align}
and $\hat{\boldsymbol{\beta}}_2^{(1)}$ be the minimizer of
\begin{align}\label{E00}
\frac{2}{n_1}\sum_{i=n_1/2+1}^{n_1}l(Y_i^{(1)},\boldsymbol X_i^{(1)\top}\boldsymbol\beta)+\gamma_{1}\|\boldsymbol\beta\|_1,
\end{align}
where $\lambda_{1}$ and $\gamma_{1}$ are two tuning parameters.
Then we store $\{\hat{\boldsymbol{\beta}}_1^{(1)}, \hat{\boldsymbol{\beta}}_2^{(1)}, n_1 \boldsymbol H_1^{(1)}, n_1\boldsymbol H_2^{(1)}\}$, where $\boldsymbol H_1^{(1)}$, and $\boldsymbol H_2^{(1)}$ are empirical versions of $\boldsymbol H$ which are obtained by using $\{\mathcal{D}_{1,1},\hat{\boldsymbol{\beta}}_2^{(1)}\}$, and $\{\mathcal{D}_{1,2},\hat{\boldsymbol{\beta}}_1^{(1)}\}$, respectively.
For
any time point $2\le s\le m$, since the raw data $\{\mathcal{D}_1,\dots\mathcal{D}_{s-1}\}$ is not stored, we consider replacing the cumulative objective function
\begin{align}
\label{E3}
&\frac{2}{N_s}\sum_{j=1}^s\sum_{i=1}^{n_j/2}l(Y_i^{(j)},\boldsymbol X_i^{(j)\top}\boldsymbol\beta)+\lambda_{s}\|\boldsymbol\beta\|_1,
\end{align}
with another function just including historical summary statistics $\{\hat{\boldsymbol\beta}_2^{(s-1)}, \sum_{j=1}^{s-1}n_j \boldsymbol H^{(j)}_1\}$, and the current dataset $\mathcal{D}_{s,1}$
 to estimate $\boldsymbol\beta^*$ at the $s$th time point, where $\lambda_{s}$ is a tuning parameter,
$N_s=\sum_{j=1}^sn_j$, $\hat{\boldsymbol\beta}^{(s-1)}_2$ is an estimator of $\boldsymbol\beta^*$ at the $(s-1)$th time point by using $\{\hat{\boldsymbol\beta}_1^{(s-2)}, \mathcal{D}_{s-1,2}, \sum_{j=1}^{s-2}n_j \boldsymbol H_2^{(j)}\}$, and $\boldsymbol H_1^{(j)}$ is an  empirical version of $\boldsymbol H$ which is acquired  by using $\{\mathcal{D}_{j,1}, \hat{\boldsymbol{\beta}}^{(j)}_2\}$
at the $j$th time point, $j=1,\dots,s-1$.
Inspiring by \cite{luo2020renewable},  
replacing $2\sum_{i=1}^{n_j/2}l(Y_i^{(j)},\boldsymbol X_i^{(j)\top}\boldsymbol\beta)/n_j$ with its second-order Taylor expansion
$(\boldsymbol\beta-\hat{\boldsymbol\beta}_2^{(s-1)})^\top  \boldsymbol H_1^{(j)}(\boldsymbol\beta-\hat{\boldsymbol\beta}_2^{(s-1)})/2+2\sum_{i=1}^{n_j/2}l(Y_i^{(j)},\boldsymbol X_i^{(j)\top}\hat{\boldsymbol\beta}_2^{(s-1)})/n_j$ in (\ref{E3}), for $j=1\dots,s-1$, and  removing constant terms, we can obtain the updating estimator
$\hat{\boldsymbol\beta}_1^{(s)}$ at the $s$th time point by minimizing the following objective function:
\begin{align}
\label{E4}
L_{1s}(\boldsymbol\beta)+\lambda_{s}\|\boldsymbol\beta\|_1,
\end{align}
where $L_{1s}(\boldsymbol\beta)=[(\boldsymbol\beta-\hat{\boldsymbol\beta}_2^{(s-1)})^\top
\sum_{j=1}^{s-1}n_j \boldsymbol H_1^{(j)}(\boldsymbol\beta-\hat{\boldsymbol\beta}_2^{(s-1)})/2+2\sum_{i=1}^{n_s/2}l(Y_i^{(s)},\boldsymbol X_i^{(s)\top}\boldsymbol\beta)]/N_s$.
Similarly, the updating estimator $\hat{\boldsymbol\beta}_2^{(s)}$ is given by
\begin{align}{\label{E111}}
\hat{\boldsymbol\beta}_2^{(s)}=\underset{\boldsymbol\beta\in\mathbb R^p}{\mbox{argmin}}\{L_{2s}(\boldsymbol\beta)+\gamma_{s}\|\boldsymbol\beta\|_1\},
\end{align}
where $L_{2s}(\boldsymbol\beta)=[(\boldsymbol\beta-\hat{\boldsymbol\beta}_1^{(s-1)})^\top
\sum_{j=1}^{s-1}n_j \boldsymbol H_2^{(j)}(\boldsymbol\beta-\hat{\boldsymbol\beta}_1^{(s-1)})/2+2\sum_{i=n_s/2+1}^{n_s}l(Y_i^{(s)},\boldsymbol X_i^{(s)\top}\boldsymbol\beta)]/N_s$,
$\gamma_{s}$ is a tuning parameter, $\hat{\boldsymbol\beta}^{(s-1)}_1$ is an estimator of $\boldsymbol\beta^*$ at the $(s-1)$th time point by using  $\{\hat{\boldsymbol\beta}_2^{(s-2)}, \mathcal{D}_{s-1,1}, \sum_{j=1}^{s-2}n_j \boldsymbol H_1^{(j)}\}$, and $\boldsymbol H_1^{(j)}$ is an  empirical version of $\boldsymbol H$ which is got by using $\{\mathcal{D}_{j,2}, \hat{\boldsymbol{\beta}}_1^{(j)}\}$
at the $j$th time point, $j=1,\dots,s-1$.
Then we take $\hat{\boldsymbol\beta}_{ave}^{(s)}=\{\hat{\boldsymbol\beta}_1^{(s)}+\hat{\boldsymbol\beta}_2^{(s)}\}/2$ as the final estimator at the $s$th step and store $\{\hat{\boldsymbol\beta}_1^{(s)},\hat{\boldsymbol\beta}_2^{(s)},\sum_{j=1}^{s}n_j \boldsymbol H_1^{(j)} ,\sum_{j=1}^{s}n_j\boldsymbol H_2^{(j)}\}$, where $\boldsymbol H_1^{(s)}$, and $\boldsymbol H_2^{(s)}$
are empirical versions of $\boldsymbol H$ which are obtained by using $\{\mathcal{D}_{s,1},\hat{\boldsymbol\beta}_2^{(s)}\}$, and $\{\mathcal{D}_{s,2},\hat{\boldsymbol\beta}_1^{(s)}\}$, respectively.  The proposed
estimation procedure is described in the following Algorithm 1.

\begin{algorithm}[htb]
{
\caption{Online estimation for the SIMs.}
\label{alg:Framwork}
\begin{algorithmic}[1] 
\renewcommand{\algorithmicrequire}{{\textbf{Input}:}}
\REQUIRE 
Streaming data sets $\mathcal{D}_1\ldots\mathcal{D}_s\ldots$, and the tuning parameters $\lambda_1\ldots\lambda_s\ldots$, $\gamma_1\ldots\gamma_s\ldots$;\\
\noindent 1: Calculate the offline lasso penalized estimators $\widehat{\bm\beta}_1^{(1)}$, $\widehat{\bm\beta}_2^{(1)}$ via (\ref{E2}) and (\ref{E00}) based on $\mathcal{D}_1$;\\
\noindent 2:~Update $n_1H_1^{(1)}$ and $n_2H_2^{(1)}$; \\
\noindent 3: \textbf{for} $s=2,3,\ldots,$ \textbf{do}\\
\noindent 4: ~~Read the current data set $\mathcal{D}_s$;\\
\noindent 5: ~~Calculate the online lasso penalized estimators $\widehat{\bm\beta}_1^{(s)}$ and $\widehat{\bm\beta}_2^{(s)}$ via (\ref{E4}) and (\ref{E111});\\
\noindent 6: ~~Update and store the summary statistics $\{\widehat{\bm\beta}_1^{(s)},\widehat{\bm\beta}_2^{(s)}, \sum_{j=1}^{s}n_j \boldsymbol H_1^{(j)}, \sum_{j=1}^{s}n_j \boldsymbol H_2^{(j)}\}$; \\
\noindent 7: ~~Calculate  $\hat{\boldsymbol\beta}_{ave}^{(s)}=\{{\hat{\boldsymbol\beta}_1}^{(s)}+{\hat{\boldsymbol\beta}_2}^{(s)}\}/2$;\\
\noindent 8: ~~Release data set $\mathcal{D}_s$ from the memory;\\
\noindent 9: \textbf{end for}\\
\renewcommand{\algorithmicrequire}{{\textbf{Output}:}}
\REQUIRE
$\widehat{\bm\beta}^{(s)}_{ave}$ for $s=1,2,\ldots$
\end{algorithmic}}
\end{algorithm}

In what follows, we will provide the convergence rates of $\hat\beta_1^{(s)}$, $\hat\beta_2^{(s)}$, and
$\hat{\boldsymbol\beta}_{ave}^{(s)}$, for $s=1,\cdots,m$.
Let $\|\cdot\|_2$ be the $\ell_2$-norm (Euclidean norm) and $\|\cdot\|_\infty$  be the maximum absolute value of the entries in a matrix. Define $N_1=n_1$,
$g_{\boldsymbol\beta}(Y,\boldsymbol  X)=\partial l(Y, \boldsymbol X^\top \boldsymbol \beta)/\partial\boldsymbol \beta$,  $\boldsymbol Z=g_{\boldsymbol\beta^*}(Y,\boldsymbol  X)$, $l_1^{(j)}(\boldsymbol\beta)=2\sum_{i=1}^{n_j/2}l(Y_i^{(j)},\boldsymbol X_i^{(j)\top}\boldsymbol\beta)/n_j$, $l_2^{(j)}(\boldsymbol\beta)=2\sum_{i=n_j/2+1}^{n_j}l(Y_i^{(j)},\boldsymbol X_i^{(j)\top}\boldsymbol\beta)/n_j$,
$\triangledown l_1^{(j)}(\boldsymbol\beta)=2\sum_{i=1}^{n_j/2}g_{\boldsymbol\beta}(Y_i^{(j)},\boldsymbol X_i^{(j)})/n_j$, and $\triangledown l_2^{(j)}(\boldsymbol\beta)=2\sum_{i=n_j/2+1}^{n_j}g_{\boldsymbol\beta}(Y_i^{(j)}, \boldsymbol X_i^{(j)})/n_j$. For a $p$-dimensional random vector $\boldsymbol\xi$, define
\begin{align*}
||\boldsymbol\xi||_{\psi_2}=\sup_{\boldsymbol a\in \mathbb R^p, ||\boldsymbol a||_2=1}\sup_{k\ge 1}(E|\boldsymbol a^\top \boldsymbol\xi|^k)^{1/k}/\sqrt{k}.
\end{align*}
In addition to condition (C1), the following conditions are required.
\begin{itemize}
\itemsep=-\parsep

\item [(C2)]
There exists a positive constant $M_1$ such that
$$
||\boldsymbol Z||_{\psi_2}\le M_1.
$$
\item [(C3)] Suppose that $\boldsymbol\beta_0$ is $s_0$-sparse with $s_0^3\log p=o(n_1^{\alpha_1})$, for some $0<\alpha_1<1$, where
$s_0$ is the number of nonzero elements in $\boldsymbol\beta_0.$

\item [(C4)]There exist two positive constant $M_2$ and $M_3$ such that
\begin{align*}
M_2\le\inf_{\|\boldsymbol\Delta\|_2=1}\|\boldsymbol H\boldsymbol\Delta\|_2
\le\sup_{\|\boldsymbol\Delta\|_2=1}\|\boldsymbol H\boldsymbol\Delta\|_2\le M_3.
\end{align*}
\item [(C5)]
 There exist two positive constants $M_4$ and $M_5$ such that for any $1\le s\le m$,
 with probability at least $1-P(n_s,p)$,
\begin{align*}
&l_1^{(s)}(\boldsymbol\beta^*+\boldsymbol\Delta)-l_1^{(s)}(\boldsymbol\beta^*)-\boldsymbol\Delta^\top \triangledown l_1^{(s)}(\boldsymbol\beta^*)\ge M_4||\boldsymbol\Delta||_2^2-M_5 \sqrt{\frac{\log p}{n_s}}
||\boldsymbol\Delta||_1||\boldsymbol\Delta||_2,\\
\mbox{and}&
\\
&l_2^{(s)}(\boldsymbol\beta^*+\boldsymbol\Delta)-l_2^{(s)}(\boldsymbol\beta^*)-\boldsymbol\Delta^\top \triangledown l_2^{(s)}(\boldsymbol\beta^*)\ge M_4||\boldsymbol\Delta||_2^2-M_5 \sqrt{\frac{\log p}{n_s}}
||\boldsymbol\Delta||_1||\boldsymbol\Delta||_2,
\end{align*}
for all $||\boldsymbol\Delta||_2\le 1$, where $\Omega(n_j,p)$ is a function of $n_j$.
\item [(C6)] There exists a positive number $M_6\ge 1$ such that for any $1\le s\le m$, with probability at least $1-P_s(n_1,\cdots,n_{s},p)$,
\begin{align*}
\max\left\{\left\|\frac{1}{N_{s}}\sum_{j=1}^{s}n_j\boldsymbol H_1^{(j)}-\boldsymbol H\right\|_\infty,\left\|\frac{1}{N_{s}}\sum_{j=1}^{s}n_j \boldsymbol H_2^{(j)}-\boldsymbol H\right\|_\infty\right\}\le \frac{1}{N_s}\sum_{j=1}^{s}n_jM_6^{2^j}\sqrt{\frac{s_0\log p}{n_j}},
\end{align*}
where $P_s(n_1,\cdots,n_{s},p)$ is a function of $n_1,\cdots,n_{s}$ and $p$.

\item [(C7)]Suppose that for any $1\le s\le m$, $2^{2^s}s_0\sqrt{\log p/N_s}=o(1)$, and
$$\lim_{p\to\infty}1-P(n_s,p)-P_{s-1}(n_1,\dots,n_{s-1},p)-2ep^{-a_0N_s/n_s}=1.$$
\end{itemize}

Condition (C2) assumes that $\boldsymbol Z$ has a sub-Gaussian tail. Condition (C3) is similar to the assumption  in
\cite{jankova2016confidence}.  Condition (C4) indicates
that $\boldsymbol H$ is positive definite and has finite eigenvalues. Conditions (C5), and (C6) are easily satisfied for many commonly-used
loss functions such as the Huber loss
\citep{huber1964robust} 
and the negative log-likelihood
of generalized linear models under some mild conditions. Condition (C7) can ensure the consistency
of our  online lasso estimators.
The following Theorem \ref{T1} provides the consistency of $\hat{\boldsymbol\beta}_1^{(s)}$, $\hat{\boldsymbol\beta}_2^{(s)}$ and $\hat{\boldsymbol\beta}_{ave}^{(s)}$, for $s=1,\cdots,m$.

\begin{theorem}
\label{T1}
Suppose that conditions (C1)-(C7) are satisfied. For any $1\le s\le m$, assume
$\lambda_{s}=c_{1s}\sqrt{\log p/N_s}$, and
$\gamma_{s}=c_{2s}\sqrt{\log p/N_s}$, where $c_{1s}$ and $c_{2s}$ could be any constants which belong to $[2M_1\sqrt{2(a_0+1)/a_1},a_2]$, $a_0$ could be any positive constant,
$a_1$ is a positive constant not depending on any parameter, and
$a_2$ could be any constant no less than $2M_1\sqrt{2(a_0+1)/a_1}$.  If
\begin{align*}
&\max_{1\le s\le m-1}a_3^{2s-2}d_1^{2^{s}}N_s^{\alpha_1/2-1/2}sM_6^{2^s}\le A_1,\\
and&\\
&\max_{1\le s\le m-1}M_5\sqrt{s_0\log p/N_{s+1}}[8+2a_3^{s-1}d_1^{2^{s-1}}M_3/\{M_1\sqrt{2(a_0+1)/a_1}\}]\le
\min\{M_2/7,M_4/3\},
\end{align*}
where $A_1$ could be any positive constant,  $d_1=\max\{12a_2/M_4,1\}$, and
\begin{align*}
a_3=[8+2M_3/\{M_1\sqrt{2(a_0+1)/a_1}\}](2M_5+3a_2/2)/\min\{M_2/3,M_4/2\}.
\end{align*}
Then for any $1\le s\le m$,
 we have that with probability at least $1-P(n_s,p)-P_{s-1}(n_1,\dots,n_{s-1},p)-2ep^{-a_0N_s/n_s}$, where $e$ is Euler’s number,
\begin{align*}
&||\hat{\boldsymbol\beta}_1^{(s)}-\boldsymbol\beta^*||_2\le a_3^{s-1}d_1^{2^{s-1}}\sqrt{\frac{s_0\log p}{N_s}}, \quad
||\hat{\boldsymbol\beta}_1^{(s)}-\boldsymbol\beta^*||_1\le a_3^{s-1}d_1^{2^{s-1}}s_0\sqrt{\frac{\log p}{N_s}},\notag \\
&||\hat{\boldsymbol\beta}_2^{(s)}-\boldsymbol\beta^*||_2\le a_3^{s-1}d_1^{2^{s-1}}\sqrt{\frac{s_0\log p}{N_s}}, \quad
||\hat{\boldsymbol\beta}_2^{(s)}-\boldsymbol\beta^*||_1\le a_3^{s-1}d_1^{2^{s-1}}s_0\sqrt{\frac{\log p}{N_s}},\notag \\
&||\hat{\boldsymbol\beta}_{ave}^{(s)}-\boldsymbol\beta^*||_2\le a_3^{s-1}d_1^{2^{s-1}}\sqrt{\frac{s_0\log p}{N_s}}, \quad
\mbox{and}
\quad ||\hat{\boldsymbol\beta}_{ave}^{(s)}-\boldsymbol\beta^*||_1\le  a_3^{s-1}d_1^{2^{s-1}}s_0\sqrt{\frac{\log p}{N_s}}.
\end{align*}
\end{theorem}
Since the proposed online estimators are developed based on
the current data batch and summary statistics of historical data, the bounds in Theorem \ref{T1} include  power functions  of $s-1$ and $2^{s-1}$, which are different from traditional oracle inequalities \citep{van2008high, huang2013oracle}.

\subsection{Online pointwise inference}
We construct pointwise inference for the $l$th
component of the regression parameter vector $\boldsymbol\beta^*$, for
$l=1,\cdots,p$. Since  $\hat{\boldsymbol\beta}_1^{(s)}, \hat{\boldsymbol\beta}_2^{(s)}$ and
$\hat{\boldsymbol\beta}_{ave}^{(s)}$ are not $N_s^{1/2}$ consistent, we cannot
obtain the asymptotic normalities of these estimators. Let $\beta_l^*$ be the $l$th element of $\boldsymbol\beta^*$, $\boldsymbol\Omega=\boldsymbol H^{-1}$, and $\hat{\boldsymbol\Omega}_1^{(s)}$
and $\hat{\boldsymbol\Omega}_2^{(s)}$ be two
estimators of $\boldsymbol\Omega$ which will be specified later.
To tackle this issue, we first consider the following one-step estimator for $\beta_l^*$ based on $\hat{\boldsymbol\beta}_1^{(s)}$ to increase the convergence rate:
\begin{align*}
\hat\beta^{one}_{1,l}=\hat\beta_{1,l}^{(s)}-\hat{\boldsymbol\Omega}_{1,l}^{(s)\top}\left\{\sum_{j=1}^{s-1}n_j\boldsymbol H_1^{(j)}(\hat{\boldsymbol\beta}_1^{(s)}-\hat{\boldsymbol\beta}_2^{(s-1)})+n_s\triangledown
l_1^{(s)}(\hat{\boldsymbol\beta}_1^{(s)})\right\}/N_s,
\end{align*}
 where $\hat\beta_{1,l}^{(s)}$ is the $l$th element of $\hat{\boldsymbol\beta}_1^{(s)}$, and
 $\hat{\boldsymbol\Omega}_{1,l}^{(s)}$ is the $l$th column of $\hat{\boldsymbol\Omega}_1^{(s)}$. It can be shown that
 \begin{align}
 \label{E5}
\hat\beta^{one}_{1,l}-\beta^*_l=&\hat\beta_{1,l}^{(s)}-\beta^*_l-\hat{\boldsymbol\Omega}_{1,l}^{(s)\top}\left\{\sum_{j=1}^{s-1}n_j\boldsymbol H_1^{(j)}(\hat{\boldsymbol\beta}_1^{(s)}-\hat{\boldsymbol\beta}_2^{(s-1)})+n_s\triangledown
l_1^{(s)}(\hat{\boldsymbol\beta}_1^{(s)})\right\}/N_s\notag \\
=&\boldsymbol\Omega_l^\top\boldsymbol H(\hat{\boldsymbol\beta}_{1}^{(s)}-\boldsymbol\beta^*)-\hat{\boldsymbol\Omega}_{1,l}^{(s)\top}\left\{\sum_{j=1}^{s-1}n_j\boldsymbol H_1^{(j)}(\hat{\boldsymbol\beta}_1^{(s)}-\hat{\boldsymbol\beta}_2^{(s-1)})+n_s\triangledown
l_1^{(s)}(\hat{\boldsymbol\beta}_1^{(s)})\right\}/N_s\notag\\
=&\boldsymbol\Omega_l^\top\sum_{j=1}^{s}n_j(\boldsymbol H-\boldsymbol H_1^{(j)})(\hat{\boldsymbol\beta}_{1}^{(s)}-\boldsymbol\beta^*)/N_s\notag \\ &-(\hat{\boldsymbol\Omega}_{1,l}^{(s)}-\boldsymbol\Omega_l)^\top\left\{\sum_{j=1}^{s-1}n_j\boldsymbol H_1^{(j)}(\hat{\boldsymbol\beta}_1^{(s)}-\hat{\boldsymbol\beta}_2^{(s-1)})+n_s\triangledown
l_1^{(s)}(\hat{\boldsymbol\beta}_1^{(s)})\right\}/N_s\notag \\
&-\boldsymbol\Omega_l^\top\left\{\sum_{j=1}^{s}n_j\boldsymbol H_1^{(j)}(\boldsymbol\beta^*-\hat{\boldsymbol\beta}_2^{(j)})+\sum_{j=1}^sn_j\triangledown
l_1^{(j)}(\hat{\boldsymbol\beta}_2^{(j)})-\sum_{j=1}^sn_j\triangledown
l_1^{(j)}(\boldsymbol\beta^*)\right\}/N_s\notag\\
&-(\boldsymbol\Omega_l-\hat{\boldsymbol\Omega}_{1,l}^{(s)})^\top\left\{\sum_{j=1}^{s-1}n_j\boldsymbol H_1^{(j)}(\hat{\boldsymbol\beta}_2^{(j)}-\hat{\boldsymbol\beta}_2^{(s-1)})-\sum_{j=1}^{s-1}n_j\triangledown
l_1^{(j)}(\hat{\boldsymbol\beta}_2^{(j)})\right\}/N_s\notag \\
&-\hat{\boldsymbol\Omega}_{1,l}^{(s)\top}\left\{\sum_{j=1}^{s-1}n_j\boldsymbol H_1^{(j)}(\hat{\boldsymbol\beta}_2^{(j)}-\hat{\boldsymbol\beta}_2^{(s-1)})-\sum_{j=1}^{s-1}n_j\triangledown
l_1^{(j)}(\hat{\boldsymbol\beta}_2^{(j)})\right\}/N_s\notag \\
&-\boldsymbol\Omega_l^\top\sum_{j=1}^{s}n_j\triangledown
l_1^{(j)}(\boldsymbol\beta^*)/N_s\notag \\
\equiv&(I)+(II)+(III)+(IV)+(V)+(VI),
\end{align}
where $\boldsymbol\Omega_l$ is the $l$th column of $\boldsymbol\Omega$. We can prove that (I)-(IV) are $o_p(N_s^{-1/2})$,  and (VI) multiply by $N_s^{-1/2}$ converges weakly to a normal distribution under some mild conditions. In addition, the order of (V) may be larger than $n^{-1/2}$. The decomposition of
$\hat\beta^{one}_{1,l}-\beta^*_l$  implies that we need to minus  (V) from (\ref{E5})  to acquire a new estimator of $\beta^*_l$ which converges weakly to a normal distribution. As a result, we propose the following estimator for $\beta_l^*$:
\begin{align}
\label{E6}
\hat\beta^{d(s)}_{1,l}=&\hat\beta^{one}_{1,l}+\hat{\boldsymbol\Omega}_{1,l}^{(s)\top}\left\{\sum_{j=1}^{s-1}n_j\boldsymbol H_1^{(j)}(\hat{\boldsymbol\beta}_2^{(j)}-\hat{\boldsymbol\beta}_2^{(s-1)})-\sum_{j=1}^{s-1}n_j\triangledown
l_1^{(j)}(\hat{\boldsymbol\beta}_2^{(j)})\right\}/N_s\notag \\
=&\hat\beta_{1,l}^{(s)}+\hat{\boldsymbol\Omega}_{1,l}^{(s)\top}\left\{\sum_{j=1}^{s-1}n_j\boldsymbol H_1^{(j)}(\hat{\boldsymbol\beta}_2^{(j)}-\hat{\boldsymbol\beta}_1^{(s)})-\sum_{j=1}^{s}n_j\triangledown
l_1^{(j)}(\hat{\boldsymbol\beta}_2^{(j)})\right\}/N_s.
\end{align}
Similarly, we propose the following estimator for $\beta_l^*$ based on $\hat{\boldsymbol\beta}_2^{(s)}$:
\begin{align}
\label{E7}
\hat\beta^{d(s)}_{2,l}=\hat\beta_{2,l}^{(s)}+\hat{\boldsymbol\Omega}_{2,l}^{(s)\top}\left\{\sum_{j=1}^{s-1}n_j\boldsymbol H_2^{(j)}(\hat{\boldsymbol\beta}_1^{(j)}-\hat{\boldsymbol\beta}_2^{(s)})-\sum_{j=1}^{s}n_j\triangledown
l_2^{(j)}(\hat{\boldsymbol\beta}_1^{(j)})\right\}/N_s,
\end{align}
 where $\hat\beta_{2,l}^{(s)}$ is the $l$th element of $\hat{\boldsymbol\beta}_2^{(s)}$, and
 $\hat{\boldsymbol\Omega}_{2,l}^{(s)}$ is the $l$th column of $\hat{\boldsymbol\Omega}_2^{(s)}$.  Subsequently,
we propose an averaged estimator to avoid efficiency loss due to sample
splitting:
\begin{align*}
\hat\beta^{da(s)}_{l}=\frac{\hat\beta^{d(s)}_{1,l}+\hat\beta^{d(s)}_{2,l}}{2}.
\end{align*}
For a matrix $\boldsymbol M\in R^{p_0\times p_1}$, let
\begin{align*}
\|\boldsymbol M\|_1=\sum_{j_1=1}^{p_0}\sum_{j_2=1}^{p_1}|M_{j_1,j_2}|,\quad
\mbox{and}\quad
\|\boldsymbol M\|_{\infty,\infty}=\max_{1\le j_2\le p_1}\sum_{j_1=1}^{p_0}|M_{j_1,j_2}|,
\end{align*}
where $M_{j_1,j_2}$ is the $(j_1,j_2)$th element of $\boldsymbol M$.
To derive upper bounds for $\|\boldsymbol \Omega-\hat{\boldsymbol\Omega}_{1}^{(s)}\|_{\infty,\infty}$ and $\|\boldsymbol \Omega-\hat{\boldsymbol\Omega}_{2}^{(s)}\|_{\infty,\infty}$ easily,
we use the method of  \cite{cai2011constrained}
to  obtain $\hat{\boldsymbol\Omega}_1^{(s)}$
and $\hat{\boldsymbol\Omega}_2^{(s)}$. For simplicity, we just present the construction of $\hat{\boldsymbol\Omega}_1^{(s)}$. Note that $\hat{\boldsymbol\Omega}_{2}^{(s)}$ can be obtained via a similar way based on $\sum_{j=1}^{s}n_j\boldsymbol H_1^{(j)}$ with the corresponding tuning parameter $\kappa_s$.
Let $\hat{\boldsymbol\Omega}$ be the solution of the following optimization problem:
\begin{align}\label{inver}
\min &\,\|\tilde{\boldsymbol\Omega}\|_1 \quad \mbox{subject to}\quad
\left\|\sum_{j=1}^{s}n_j\boldsymbol H_1^{(j)}\tilde{\boldsymbol\Omega}/N_s-\boldsymbol I_p\right\|_\infty\le h_s, 
\end{align}
where $h_s$ is a tuning parameter and $\boldsymbol I_p$ is a unit matrix of size $p$.
Note that the solution of (\ref{inver}) is not symmetric in general. The final  estimator $\hat{\boldsymbol\Omega}_1^{(s)}$ is obtained by symmetrizing $\hat{\boldsymbol\Omega}$ as follows:
\begin{align}
\hat\Omega_{1,j_1,j_2}^{(s)}=\hat\Omega_{1, j_2,j_1}^{(s)}=\hat\Omega_{j_1,j_2}I(|\hat\Omega_{j_1,j_2}|\le|\hat\Omega_{j_2,j_1}|)
+\hat\Omega_{j_2,j_1}I(|\hat\Omega_{j_2,j_1}|<|\hat\Omega_{j_1,j_2}|),\notag
\end{align}
where $\hat\Omega_{1,j_1,j_2}^{(s)}$, and $\hat\Omega_{j_1,j_2}$ are the $(j_1,j_2)$th elements
of $\hat{\boldsymbol\Omega}_{1}^{(s)}$ and
$\hat{\boldsymbol\Omega}$, respectively, and $\hat\Omega_{1,j_2,j_1}^{(s)}$, and $\hat\Omega_{j_2,j_1}$ are the $(j_2,j_1)$th elements
of $\hat{\boldsymbol\Omega}_{1}^{(s)}$, and $\hat{\boldsymbol\Omega}$, respectively.
  Both (\ref{E6}) and (\ref{E7}) imply that
 $\{\sum_{j=1}^{s-1}n_j\boldsymbol H_1^{(j)}\hat{\boldsymbol\beta}_2^{(j)}-\sum_{j=1}^{s-1}n_j\triangledown
l_1^{(j)}(\hat{\boldsymbol\beta}_2^{(j)})\}$ and $\{\sum_{j=1}^{s-1}n_j\boldsymbol H_2^{(j)}\hat{\boldsymbol\beta}_1^{(j)}-\sum_{j=1}^{s-1}n_j\triangledown
l_2^{(j)}(\hat{\boldsymbol\beta}_1^{(j)})\}$ should be stored as historical summary statistics at the $(s-1)$th step
to acquire $\hat\beta^{d(s)}_{1,l}$ and $\hat\beta^{d(s)}_{2,l}$. In addition, we should also store $T_s$, which
is defined as
\begin{align*}
T_s=&\frac{1}{N_s}\Bigg\{\sum_{j=1}^s\sum_{i=1}^{n_j/2}g_{\hat{\boldsymbol\beta}_2^{(j)}}(Y_i^{(j)},\boldsymbol  X_i^{(j)})g^\top_{\hat{\boldsymbol\beta}_2^{(j)}}(Y_i^{(j)},\boldsymbol X_i^{(j)})\\
&+\sum_{j=1}^s\sum_{i=n_j/2+1}^{n_j}g_{\hat{\boldsymbol\beta}_1^{(j)}}(Y_i^{(j)},\boldsymbol  X_i^{(j)})g^\top_{\hat{\boldsymbol\beta}_1^{(j)}}(Y_i^{(j)},\boldsymbol X_i^{(j)})\Bigg\}
\end{align*}
to estimate the asymptotic variance of $\sqrt{N_s}(\hat\beta^{da(s)}_{l}-\beta_l^*)$. Denote $Q_1^{(s-1)} = \sum_{j=1}^{s-1}n_j\boldsymbol H_1^{(j)}\hat{\boldsymbol\beta}_2^{(j)}-\sum_{j=1}^{s-1}n_j\triangledown
l_1^{(j)}(\hat{\boldsymbol\beta}_2^{(j)})$ and $Q_2^{(s-1)}=\sum_{j=1}^{s-1}n_j\boldsymbol H_2^{(j)}\hat{\boldsymbol\beta}_1^{(j)}-\sum_{j=1}^{s-1}n_j\triangledown
l_2^{(j)}(\hat{\boldsymbol\beta}_1^{(j)}).$
The proposed debiasing procedure is presented in the following Algorithm 2.

\begin{algorithm}[htb]
{
\caption{Online pointwise inference for the SIMs.}
\label{alg:Framwork2}
\begin{algorithmic}[1] 
\renewcommand{\algorithmicrequire}{{\textbf{Input}:}}
\REQUIRE 
Streaming data sets $\mathcal{D}_1\ldots\mathcal{D}_s\ldots$;\\
\noindent 1: Calculate the offline lasso penalized estimators $\widehat{\bm\beta}_1^{(1)}$, $\widehat{\bm\beta}_2^{(1)}$ via (\ref{E2}) and (\ref{E00}) based on $\mathcal{D}_1$;\\
\noindent 2: Update $n_1H_1^{(1)}$, $n_1H_2^{(1)}$, $Q_1^{(1)}$, $Q_2^{(1)}$ and $T_1$; \\
\noindent 3: \textbf{for} $s=2,3,\ldots,$ \textbf{do}\\
\noindent 4: ~~Read the current data set $\mathcal{D}_s$;\\
\noindent 5: ~~Update online lasso penalized estimators $\widehat{\bm\beta}^{(s)}_1$  and $\widehat{\bm\beta}^{(s)}_2$ via Algorithm 1;\\
\noindent 6: ~~Update and store the summary statistics $\{\sum_{j=1}^{s}n_j\boldsymbol H_1^{(j)},\sum_{j=1}^{s}n_j\boldsymbol H_2^{(j)}, Q_1^{(s)}, Q_2^{(s)},T_s\}$;\\
\noindent 7: ~~Calculate $\hat{\boldsymbol\Omega}_{1}^{(s)}$ and $\hat{\boldsymbol\Omega}_{2}^{(s)}$ by using (\ref{inver});\\
\noindent 8: ~~Update the online debiasing estimators $\hat\beta^{d(s)}_{1,l}$ and $\hat\beta^{d(s)}_{2,l}$ via (\ref{E6}) and (\ref{E7});\\
\noindent 9: ~~Compute  $\hat{\boldsymbol\beta}_{l}^{da(s)}=\{{\hat{\boldsymbol\beta}_{1,l}}^{da(s)}+{\hat{\boldsymbol\beta}_{2,l}}^{da(s)}\}/2$  and $\hat\sigma_{l,s}^{2}$ by (\ref{varian});\\
\noindent 10: ~Release data set $\mathcal{D}_s$ from the memory;\\
\noindent 11: \textbf{end for}\\ %
\renewcommand{\algorithmicrequire}{{\textbf{Output}:}}
\REQUIRE
$\hat{\boldsymbol\beta}_{l}^{da(s)}$ and $\hat\sigma_{l,s}^{2}$ for $s=1,2,\ldots$
\end{algorithmic}}
\end{algorithm}

Let $\sigma_{l}^2=\boldsymbol{\Omega}_l^\top E(\boldsymbol Z\boldsymbol {Z}^\top)\boldsymbol{\Omega}_l$.
Additional conditions are needed to prove Theorem \ref{T2}.
\begin{itemize}
\itemsep=-\parsep

\item [(D1)] Assume that for any $1\le l\le p$,
$$
\sigma_{l}^2\ge G_1,
$$
where $G_1$ is a positive constant.
\item [(D2)] There exists a positive number
$v(p)$ depending on $p$, and
  a positive constant $\omega$ which belongs to $[0,1)$ such that for any $1\le s\le m$,
\begin{align*}
\max\{\|\hat{\boldsymbol\Omega}_1^{(s)}-\boldsymbol\Omega\|_{\infty,\infty},
\|\hat{\boldsymbol\Omega}_2^{(s)}-\boldsymbol\Omega\|_{\infty,\infty}\}=O_p((g(s,s_0)\|\boldsymbol\Omega\|_{\infty,\infty}^4\log p/N_s)^{(1-\omega)/2}v(p)),
\end{align*}
where $g(s,s_0)$ is a function of $s$ and $s_0$.
\item [(D3)]For any $1\le s\le m$,
assume
\begin{align*}
&\|\boldsymbol\Omega\|_{\infty,\infty}\left\|\left\{\sum_{j=1}^{s}n_j\boldsymbol H_1^{(j)}(\boldsymbol\beta^*-\hat{\boldsymbol\beta}_2^{(j)})+\sum_{j=1}^sn_j\triangledown
l_1^{(j)}(\hat{\boldsymbol\beta}_2^{(j)})-\sum_{j=1}^sn_j\triangledown
l_1^{(j)}(\boldsymbol\beta^*)\right\}/N_s^{1/2}\right\|_\infty=o_p(1),\\
&\mbox{and}\\
&\|\boldsymbol\Omega\|_{\infty,\infty}\left\|\left\{\sum_{j=1}^{s}n_j\boldsymbol H_2^{(j)}(\boldsymbol\beta^*-\hat{\boldsymbol\beta}_1^{(j)})+\sum_{j=1}^sn_j\triangledown
l_2^{(j)}(\hat{\boldsymbol\beta}_1^{(j)})-\sum_{j=1}^sn_j\triangledown
l_2^{(j)}(\boldsymbol\beta^*)\right\}/N_s^{1/2}\right\|_\infty=o_p(1).
\end{align*}
\item [(D4)] Assume
\begin{align*}
&\{g(s,s_0)\}^{(1-\omega)/2}\|\boldsymbol\Omega\|_{\infty,\infty}^{2(1-\omega)}v(p)(\log p)^{1-\omega/2}/n_1^{(1-\omega)/2}=o(1),\\
&\|\boldsymbol\Omega\|_{\infty,\infty} \max_{1\le s\le m} a_3^{s-1}d_1^{2^{s-1}}N_s^{\alpha_1/2-1/2}s\sqrt{\log p}M_6^{2^s}\le A_1,\\
\mbox{and}&\\
&\{g(s,s_0)\}^{(1-\omega)/2}\|\boldsymbol\Omega\|_{\infty,\infty}^{2(1-\omega)}s_0\max_{1\le s\le m}(s-1)v(p)(\log p)^{1-\omega/2} a_3^{s-2}d_1^{2^{s-2}}N_s^{\alpha_1/2+\omega/2-1}M_6^{2^{s-1}}\le A_1,
\end{align*}
and for any $2\le s\le m$,
\begin{align*}
&\{g(s,s_0)\}^{(1-\omega)/2}\|\boldsymbol\Omega\|_{\infty,\infty}^{2(1-\omega)}a_3^{s-2}d_1^{2^{s-2}}s_0(\log p)^{1-\omega/2}v(p)N_s^{\omega/2-1/2}=o(1).
\end{align*}
\end{itemize}
Condition (D1) assumes that the
 asymptotic variance of $\sqrt{N_s}(\hat\beta^{da(s)}_{l}-\beta_l^*)$ is bounded away from zero.
 $\max\{\|\hat{\boldsymbol\Omega}_1^{(s)}-\boldsymbol\Omega\|_{\infty,\infty},
\|\hat{\boldsymbol\Omega}_2^{(s)}-\boldsymbol\Omega\|_{\infty,\infty}\}=O_p((g(s,s_0)\|\boldsymbol\Omega\|_{\infty,\infty}^4\log p/N_s)^{(1-\omega)/2}v(p))$, and
condition (D3) are easily satisfied under some mild conditions. Conditions (D2)-(D4) can ensure that the first four terms on the right side of (\ref{E5}) are $o_p(N_s^{-1/2})$.
The following theorem  demonstrates the asymptotic properties
of $\sqrt{N_s}(\hat\beta^{da(s)}_{l}-\beta_l^*)$.
\begin{theorem}
\label{T2}
Under the conditions of Theorem \ref{T1}, suppose
that conditions (D1)-(D4) are satisfied. Then for any $1\le s\le m$ and  $1\le l\le p$,
 we have $\sigma_{l}^{-1}\sqrt{N_s}(\hat\beta^{da(s)}_{l}-\beta_l^*)\longrightarrow N(0,1)$ in distribution as $p\to \infty$.
\end{theorem}
 The asymptotic variance of $\sqrt{N_s}(\hat\beta^{da(s)}_{l}-\beta_l^*)$
can be estimated by
\begin{align}{\label{varian}}
\hat\sigma_{l,s}^{2}=&(\hat{\boldsymbol\Omega}_{1,l}^{(s)}+ \hat{\boldsymbol\Omega}_{2,l}^{(s)})^\top
T_s(\hat{\boldsymbol\Omega}_{2,l}^{(s)}+ \hat{\boldsymbol\Omega}_{2,l}^{(s)})/4.
\end{align}
Then for any given significant
level $\alpha\in(0,1)$, a $(1-\alpha)$ confidence interval for $\beta_l^*$ is
\begin{align*}
[\hat\beta^{da(s)}_{l}-N_s^{-1/2}\hat\sigma_{l,s}z_{\alpha/2},\hat\beta^{da(s)}_{l}+N_s^{-1/2}\hat\sigma_{l,s}z_{\alpha/2}],
\end{align*}
where $z_{\alpha/2}$ is the upper $\alpha/2$-quantile of the standard normal distribution.

\section{Examples}

In this section, we provide two concrete examples to
illustrate the proposed method.

\subsection{Huber loss}

Actually, we often encounter data subject to heavily-tailed errors in finance and economics \citep{fan2017estimation,fan2021shrinkage}. The Huber loss as an important way of robustification has been well studied recently \citep{fan2017estimation,sun2020adaptive,loh2021scale,wang2021new}.  The Huber loss function is defined as follows:
$$l(Y,\boldsymbol X^\top\boldsymbol\beta)=\rho_\tau(Y-\boldsymbol X^\top\boldsymbol\beta),$$ where
\begin{align*}
\rho_\tau(x)=\frac{x^2}{2}I(|x|\le\tau)+(\tau|x|-\frac{\tau^2}{2})
I(|x|>\tau),
\end{align*}
for some constant $\tau>0.$ We can observe that the Huber loss is robust to the heavy-tailed observation noise due to the fact that the linear part of the Huber loss penalizes the residuals. Let $\boldsymbol\beta_\tau^*=\mbox{argmin}_{\boldsymbol\beta\in\mathbb{R}^p}E\{\rho_\tau(Y-\boldsymbol X^\top\boldsymbol\beta)\}$, and
$\epsilon_\tau=Y-\boldsymbol X^\top\boldsymbol \beta_\tau^*$. If $\epsilon_\tau$ is a continuous random variable, then we have
\begin{align*}
&\boldsymbol H_\tau=\frac{\partial^2 }{\partial{\boldsymbol\beta}
\partial{\boldsymbol\beta}^\top}E\{\rho_\tau(Y-\boldsymbol X^\top\boldsymbol\beta)\}|_{\boldsymbol\beta=\boldsymbol\beta_\tau^*}
=E\{\boldsymbol X\boldsymbol X^\top I(|\epsilon_\tau|\le\tau)\},\\
&\boldsymbol H_1^{(s)}=\frac{2}{n_s}\sum_{i=1}^{n_s/2}\boldsymbol X_i^{(s)}\boldsymbol X_i^{(s)\top } I(|Y_i^{(s)}-\boldsymbol X_i^{(s)\top}\hat{\boldsymbol\beta}_2^{(s)}|\le\tau),\\
\mbox{and}&\\
&\boldsymbol H_2^{(s)}=\frac{2}{n_s}\sum_{i=n_s/2+1}^{n_s}\boldsymbol X_i^{(s)}\boldsymbol X_i^{(s)\top } I(|Y_i^{(s)}-\boldsymbol X_i^{(s)\top}\hat{\boldsymbol\beta}_1^{(s)}|\le\tau),\quad s=1,\cdots,m.
\end{align*}
We can obtain the estimators $\hat{\boldsymbol\beta}_1^{(s)}$, $\hat{\boldsymbol\beta}_2^{(s)}$
 and $\hat{\boldsymbol\beta}_{ave}^{(s)}$ by using the estimation procedure in Algorithm 1, for $s=1,\cdots,m$.

The following conditions are needed to establish the consistency of $\hat{\boldsymbol\beta}_1^{(s)}$, $\hat{\boldsymbol\beta}_2^{(s)}$
 and $\hat{\boldsymbol\beta}_{ave}^{(s)}$.
\begin{itemize}
\itemsep=-\parsep
\item [(E1)]   Assume that $\boldsymbol X$ is independent of $\epsilon$, and satisfies the linearity of expectation assumption in the direction of $\boldsymbol\beta_0$. In addition, there exists a positive constant $e_1$ such that for any $\tau>e_1$, the function $\boldsymbol\beta\rightarrow E\{\rho_{\tau}(Y-\boldsymbol X^\top\boldsymbol\beta)\}$ has a unique minimizer $\boldsymbol\beta^*_\tau\ne 0$.
\item [(E2)] There exists a positive constant $B_1$ such that $\|\boldsymbol X\|_{\psi_2}\le B_1$.
\item [(E3)]  There exist two positive constants  $B_2$ and $B_3$ such that for any $\tau>e_1$,
$E|\epsilon_\tau|\le B_2$, and
\begin{align*}
 B_3\le\inf\limits_{\|\boldsymbol\Delta\|_2=1}
 {\|\boldsymbol H_\tau^{1/2}\boldsymbol\Delta\|_2^2}
 \le
\sup\limits_{\|\boldsymbol\Delta\|_2=1}{\|\boldsymbol H_\tau^{1/2}\boldsymbol\Delta\|_2^2}\le B_2.
\end{align*}
\item [(E4)]  There exist two positive constants  $B_{4}$ and $0<\alpha_2<1$ such that for any $2\le s\le m$,
\begin{align*}
 \frac{\log p}{n_s}\le B_{4} \quad\mbox{or}\quad \log p/n_s> (\log p)^{\alpha_2}.
\end{align*}
\item [(E5)] For any given $\tau>e_1$, there exists a positive constant $L_\tau$ depending on $\tau$ such that  $\mathop{}\sup\limits_{{x\in R}}f_{\epsilon_\tau|X}(x) \le L_\tau$ almost surely, where
    $f_{\epsilon_\tau|\boldsymbol X}(\cdot)$ is the conditional density function of $\epsilon_\tau$ given $\boldsymbol X$.
\end{itemize}

Condition (E1), which is similar to condition (C1), is for the parameter identification. Condition (E2) implies condition (C2). Conditions (E2)-(E4) imply condition (C5). Conditions (E2)-(E5) imply condition (C6).  The
following Corollary \ref{C1}  provide the $\ell_1$  and $\ell_2$ bounds for $\hat{\boldsymbol\beta}_1^{(s)}$, $\hat{\boldsymbol\beta}_2^{(s)}$
 and $\hat{\boldsymbol\beta}_{ave}^{(s)}$.

\begin{corollary}
\label{C1}
Suppose that conditions (C3)  and (E1)-(E5) hold. For any $1\le s\le m$, assume
$\lambda_{s}=c_{1s}'\sqrt{\log p/N_s}$ and
$\gamma_{s}=c_{2s}'\sqrt{\log p/N_s}$, where $c_{1s}'$ and $c_{2s}'$ could be any constants which belong to $[2\tau M_7\sqrt{2(a_0'+1)/a_1},a_2']$, $a_0'$ could be any positive constant, and
$a_2'$ could be any constant no less than $2\tau B_1\sqrt{2(a_0'+1)/a_1}$. If $\tau\ge g_1$,
\begin{align*}
&\max_{1\le s\le m-1}a_3'^{2s-2}d_1'^{2^{2s-2}}N_s^{\alpha_1/2-1/2}sM_\tau^{2^s}\le A_1',\\
&\max_{1\le s\le m-1}g_3\sqrt{s_0\log p/N_{s+1}}[8+2a_3'^{s-1}d_1'^{2^{s-1}}B_2/\{\tau B_1\sqrt{2(a_0'+1)/a_1}\}]\le
\min\{B_3/7,g_2/3\},\\
&a_3'=[8+2M_2/\{\tau B_1\sqrt{2(a_0'+1)/a_1}\}](2B_2+3a_2'/2)/\min\{B_3/3,g_2/2\},\\
\mbox{and}&\\
&M_\tau^2=[\max\{\sqrt{32B_1^4(a_0'+2)/a_4'},8B_1^2(a_0'+2)/a_4'\}+
 4\sqrt{2}L_\tau B_1^3+1] a_3'd_1',
\end{align*}
where $A_1'$ could be any constant,  $d_1'=\max\{12a_2'/g_2,1\}$, $a_4'$ is a positive constant not depending on any parameter, and $g_1$, $g_2$  and $g_3$ are three positive constants depending on $e_1$, $B_1$, $B_2$, $B_3$  and $B_{4}$.
Then for any $1\le s\le m$,
 we have that with probability at least $1-4(s-1)p^{-a_0'}-\sum_{j=1}^{s}\{\exp(-g_4n_j-g_5\log p)+2ep^{-a_0'N_{j}/n_{j}}\}$,
\begin{align*}
&||\hat{\boldsymbol\beta}_1^{(s)}-\boldsymbol\beta_\tau^*||_2\le a_3'^{s-1}d_1'^{2^{s-1}}\sqrt{\frac{s_0\log p}{N_s}}, \quad
||\hat{\boldsymbol\beta}_1^{(s)}-\boldsymbol\beta_\tau^*||_1\le a_3'^{s-1}d_1'^{2^{s-1}}s_0\sqrt{\frac{\log p}{N_s}},\notag \\
&||\hat{\boldsymbol\beta}_2^{(s)}-\boldsymbol\beta_\tau^*||_2\le a_3'^{s-1}d_1'^{2^{s-1}}\sqrt{\frac{s_0\log p}{N_s}}, \quad
||\hat{\boldsymbol\beta}_2^{(s)}-\boldsymbol\beta_\tau^*||_1\le a_3'^{s-1}d_1'^{2^{s-1}}s_0\sqrt{\frac{\log p}{N_s}},\notag \\
&||\hat{\boldsymbol\beta}_{ave}^{(s)}-\boldsymbol\beta_\tau^*||_2\le a_3'^{s-1}d_1'^{2^{s-1}}\sqrt{\frac{s_0\log p}{N_s}}, \quad
\mbox{and}
\quad ||\hat{\boldsymbol\beta}_{ave}^{(s)}-\boldsymbol\beta_\tau^*||_1\le  a_3'^{s-1}d_1'^{2^{s-1}}s_0\sqrt{\frac{\log p}{N_s}},
\end{align*}
where $g_4$  and $g_5$ are two positive constants depending on $e_1$, $B_1$, $B_2$, $B_3$  and $B_{4}$.
\end{corollary}
When $m=o(\min(p^{-a_0'},p^{-g_5}))$, the $\ell_1$ and $\ell_2$ norms between
the estimators $\hat{\boldsymbol\beta}_1^{(s)}, \hat{\boldsymbol\beta}_2^{(s)}$, and
$\hat{\boldsymbol\beta}_{ave}^{(s)}$
 and $\beta_\tau^*$ in Corollary \ref{C1} are of orders $\sqrt{s_0^2\log p/(M_\tau^{2^s}sN_s^{\alpha_1/2+1/2})}$ and
 $\sqrt{s_0\log p/(M_\tau^{2^s}sN_s^{\alpha_1/2+1/2})}$, respectively.

The following conditions are required for the asymptotic normality of $\hat\beta^{da(s)}_{l}$.

\begin{itemize}
\itemsep=-\parsep

\item [(E6)]
There exist a constant $G_1'$ such that for any $\tau\ge e_1$  and   $1\le l\le p$,
$$
\sigma_{\tau, l}^2\ge G_1'.
$$
\item [(E7)] Suppose that for any $\tau\ge e_1$,
\begin{align}
\max_{1\le j\le p} \sum_{k=1}^{p}|\Omega_{\tau,k,j}|^{\omega}\le v(p),\notag
\end{align}
where $\Omega_{\tau,k,j}$ is the $(k,j)$th element of $\boldsymbol\Omega_\tau$.
\item [(E8)] Assume that for any $\tau\ge e_1$,
\begin{align*}
&m=o(\min(p^{-a_0'},p^{-g_5})),\\
&\{s^2M_\tau^{2^{s+1}}s_0\}^{(1-\omega)/2}\|\boldsymbol\Omega_\tau\|_{\infty,\infty}^{2(1-\omega)}v(p)(\log p)^{1-\omega/2}/n_1^{(1-\omega)/2}=o(1),\\
&\|\boldsymbol\Omega_\tau\|_{\infty,\infty}\max_{1\le s\le m}a_3'^{s-1}d_1'^{2^{s-1}}N_s^{\alpha_1/2-1/2}s\sqrt{\log p}M_\tau^{2^s}\le A_1',\\
&\|\boldsymbol\Omega_\tau\|_{\infty,\infty}\max_{1\le s\le m}a_3'^{s-1}d_1'^{2^{s-1}}s_0^{1/2}N_s^{-1/2}\log p=o(1),\\
&\{s^2M_\tau^{2^{s+1}}s_0\}^{(1-\omega)/2}\|\boldsymbol\Omega_\tau\|_{\infty,\infty}^{2(1-\omega)}s_0\max_{1\le s\le m}(s-1)v(p)(\log p)^{1-\omega/2}a_3'^{s-2}d_1'^{2^{s-2}}N_s^{\alpha_1/2+\omega/2-1}M_\tau^{2^{s-1}}\le A_1',
\end{align*}
for any $2\le s\le m$,
\begin{align*}
&\{s^2M_\tau^{2^{s+1}}s_0\}^{(1-\omega)/2}\|\boldsymbol\Omega_\tau\|_{\infty,\infty}^{2(1-\omega)}a_3'^{s-2}d_1'^{2^{s-2}}s_0(\log p)^{1-\omega/2}v(p)N_s^{\omega/2-1/2}=o(1),\\
\end{align*}
and for any $1\le s\le m$,
\begin{align*}
&\|\boldsymbol\Omega_\tau\|_{\infty,\infty}a_3'^{2s-2}d_1'^{2^{s}}sN_s^{\alpha_1-1/2}=o(1).
\end{align*}

\end{itemize}
Condition (E6) is similar to condition (D1).
Condition (E7) is for deriving the upper bound for $\max\{\|\hat{\boldsymbol\Omega}_1^{(s)}-\boldsymbol\Omega_\tau\|_{\infty,\infty},
\|\hat{\boldsymbol\Omega}_2^{(s)}-\boldsymbol\Omega_\tau\|_{\infty,\infty}\}$
\citep{cai2011constrained}. 
Condition (E8) is similar to
condition (D4). In condition (E8), $g(s,s_0)=s^2M_\tau^{2^{s+1}}s_0$.
The following corollary provides the asymptotic distribution of $\sqrt{N_s}(\hat\beta^{da(s)}_{l}-\beta_{\tau,l}^*)$, where $\beta_{\tau,l}^*$ is the
$l$th element of  $\boldsymbol\beta_{\tau}^*$.

\begin{corollary}
\label{C2}
Under the same conditions of Corollary \ref{C1}, suppose in addition
that conditions (E6)-(E8) are satisfied and for any $1\le s\le m$,
$h_s=c_{3s}'sM_\tau^{2^s}s_0^{1/2}\|\boldsymbol\Omega_\tau\|_{\infty,\infty}\sqrt{\log p/N_s}$  and $\kappa_s=c_{4s}'sM_\tau^{2^s}s_0^{1/2}\|\boldsymbol\Omega_\tau\|_{\infty,\infty}\sqrt{\log p/N_s}$, where $c_{3s}'$  and $c_{4s}'$ could be any constants no less than $1$. Then for any $1\le s\le m$ and $1\le l\le p$,
 we have $\sigma_{\tau, l}^{-1}\sqrt{N_s}(\hat\beta^{da(s)}_{l}-\beta_{\tau, l}^*)\longrightarrow N(0,1)$ in distribution as $p\to \infty$.
\end{corollary}

\subsection{Logistic loss}

{
 If $Y$ is a binary outcomes that takes only the value $0$ or $1$,
 the logistic regression models are widely used in finance, business, computer
science, and genetics \citep{hosmer2013applied,sur2019modern,ma2021global}.
In this example, we consider the following negative log-likelihood  as the loss function:
 }
\begin{align*}
l(Y,\boldsymbol X^\top\boldsymbol\beta)
=\log\{1+\exp(\boldsymbol X^\top\boldsymbol\beta)\}-Y\boldsymbol X^\top\boldsymbol\beta.
\end{align*}
Then we have
\begin{align*}
&\boldsymbol H=\frac{\partial^2 }{\partial{\boldsymbol\beta}
\partial{\boldsymbol\beta}^\top}E\{l(Y-\boldsymbol X^\top\boldsymbol\beta)\}|_{\boldsymbol\beta=\boldsymbol\beta^*}
=E[\boldsymbol X\boldsymbol X^\top \frac{\exp(\boldsymbol X^\top\boldsymbol\beta)}
{\{1+\exp(\boldsymbol X^\top\boldsymbol\beta)\}^2}],\\
&\boldsymbol H_1^{(s)}=\frac{2}{n_s}\sum_{i=1}^{n_s/2}\boldsymbol X_i^{(s)}\boldsymbol X_i^{(s)\top }
\frac{\exp(\boldsymbol X_i^{(s)\top}\hat{\boldsymbol\beta}_2^{(s)})}
{\{1+\exp(\boldsymbol X_i^{(s)\top}\hat{\boldsymbol\beta}_2^{(s)})\}^2},\\
\mbox{and}&\\
&\boldsymbol H_2^{(s)}=\frac{2}{n_s}\sum_{i=n_s/2+1}^{n_s}\boldsymbol X_i^{(s)}\boldsymbol X_i^{(s)\top }
\frac{\exp(\boldsymbol X_i^{(s)\top}\hat{\boldsymbol\beta}_1^{(s)})}
{\{1+\exp(\boldsymbol X_i^{(s)\top}\hat{\boldsymbol\beta}_1^{(s)})\}^2},\quad s=1,\cdots,m.
\end{align*}
The following Corollary \ref{C3} presents the consistency of $\hat{\boldsymbol\beta}_1^{(s)}$, $\hat{\boldsymbol\beta}_2^{(s)}$
 and $\hat{\boldsymbol\beta}_{ave}^{(s)}$.
\begin{corollary}
\label{C3}
Assume that conditions (C1), (C3), (C4), (E2)  and (E4) are satisfied. For any $1\le s\le m$, assume
$\lambda_{s}=c_{1s}''\sqrt{\log p/N_s}$  and
$\gamma_{s}=c_{2s}''\sqrt{\log p/N_s}$, where $c_{1s}''$  and $c_{2s}''$ could be any constants which belong to $[2B_1\sqrt{2(a_0''+1)/a_1},a_2'']$, $a_0''$ could be any positive constant, and
$a_2''$ could be any constant no less than $2B_1\sqrt{2(a_0''+1)/a_1}$. If
\begin{align*}
&\max_{1\le s\le m-1}a_3''^{2s-2}d_1''^{2^{s}}N_s^{\alpha_1/2-1/2}s\tilde M^{2^s} \le A_1'', \notag\\
&\max_{1\le s\le m-1}g_2'\sqrt{s_0\log p/N_{s+1}}[8+2a_3''^{s-1}d_1''^{2^{s-1}}M_3/\{ B_1\sqrt{2(a_0''+1)/a_1}\}]\le
\min\{M_2/7,g_1'/3\},\\
&a_3''=[8+2M_3/\{B_1\sqrt{2(a_0'+1)/a_1}\}](2M_3+3a_2''/2)/\min\{M_2/3,g_1'/2\},\\
\mbox{and} &\\
&\tilde M^2=[\max\{\sqrt{32B_1^4(a_0''+2)/a_4'},8B_1^2(a_0''+2)/a_4'\}+
 4\sqrt{2} B_1^3+1] a_3''d_1'',
\end{align*}
where $A_1''$ could be any constant, $d_{s}''=a_3''^{s-1}d_1''^{2^{s-1}}$, $d_1''=\max\{12a_2''/g_1',1\}$, and $g_1'$ and $g_2'$ are two positive constants depending on $M_2$, $M_3$, $B_1$,  and $B_{4}$.
Then for
any $1\le s\le m$,
 we have that with probability at least $1-4(s-1)p^{-a_0''}-\sum_{j=1}^{s}\{\exp(-g_3'n_j-g_4'\log p)+2ep^{-a_0''N_{j}/n_{j}}\}$,
\begin{align*}
&||\hat{\boldsymbol\beta}_1^{(s)}-\boldsymbol\beta^*||_2\le a_3''^{s-1}d_1''^{2^{s-1}}\sqrt{\frac{s_0\log p}{N_s}}, \quad
||\hat{\boldsymbol\beta}_1^{(s)}-\boldsymbol\beta^*||_1\le a_3''^{s-1}d_1''^{2^{s-1}}s_0\sqrt{\frac{\log p}{N_s}},\notag \\
&||\hat{\boldsymbol\beta}_2^{(s)}-\boldsymbol\beta^*||_2\le a_3''^{s-1}d_1''^{2^{s-1}}\sqrt{\frac{s_0\log p}{N_s}}, \quad
||\hat{\boldsymbol\beta}_2^{(s)}-\boldsymbol\beta^*||_1\le a_3''^{s-1}d_1''^{2^{s-1}}s_0\sqrt{\frac{\log p}{N_s}},\notag \\
&||\hat{\boldsymbol\beta}_{ave}^{(s)}-\boldsymbol\beta_\tau^*||_2\le a_3''^{s-1}d_1''^{2^{s-1}}\sqrt{\frac{s_0\log p}{N_s}}, \quad
\mbox{and}
\quad ||\hat{\boldsymbol\beta}_{ave}^{(s)}-\boldsymbol\beta^*||_1\le  a_3''^{s-1}d_1''^{2^{s-1}}s_0\sqrt{\frac{\log p}{N_s}},
\end{align*}
where $g_3'$  and $g_4'$ are two positive constants depending on $M_2$, $M_3$, $B_1$   and $B_{4}$.
\end{corollary}
If $m=o(\min(p^{-a_0''},p^{-g_4'}))$, then the $\ell_1$ and $\ell_2$ norms between
the estimators in Corollary \ref{C2}
 and $\beta^*$  are of orders $\sqrt{s_0^2\log p/(\tilde M^{2^s}sN_s^{\alpha_1/2+1/2})}$ and
 $\sqrt{s_0\log p/(\tilde M^{2^s}sN_s^{\alpha_1/2+1/2})}$, respectively.

 Two additional  conditions are needed to prove the asymptotic normality
of $\hat\beta^{da(s)}_{l}$.
\begin{itemize}
\itemsep=-\parsep
\item [(E9)] Assume
\begin{align}
\max_{1\le j\le p} \sum_{k=1}^{p}|\Omega_{k,j}|^{\omega}\le v(p).\notag
\end{align}
\item [(E10)] Assume
\begin{align*}
&m=o(\min(p^{-a_0''},p^{-g_4'})),\\
&\{s^2\tilde M^{2^{s+1}}s_0\}^{(1-\omega)/2}\|\boldsymbol\Omega\|_{\infty,\infty}^{2(1-\omega)}v(p)(\log p)^{1-\omega/2}/n_1^{(1-\omega)/2}=o(1),\\
&\|\boldsymbol\Omega\|_{\infty,\infty}\max_{1\le s\le m}a_3''^{s-1}d_1''^{2^{s-1}}N_s^{\alpha_1/2-1/2}s\sqrt{\log p}\tilde M^{2^s}\le A_1'',\\
&\|\boldsymbol\Omega\|_{\infty,\infty}\max_{1\le s\le m}a_3''^{s-1}d_1''^{2^{s-1}}s_0^{1/2}N_s^{-1/2}\log p=o(1),\\
&\{s^2\tilde M^{2^{s+1}}s_0\}^{(1-\omega)/2}\|\boldsymbol\Omega\|_{\infty,\infty}^{2(1-\omega)}s_0\max_{1\le s\le m}(s-1)v(p)(\log p)^{1-\omega/2}a_3''^{s-2}d_1''^{2^{s-2}}N_s^{\alpha_1/2+\omega/2-1}\tilde M^{2^{s-1}}\le A_1'',
\end{align*}
for any $2\le s\le m$,
\begin{align*}
&\{s^2\tilde M^{2^{s+1}}s_0\}^{(1-\omega)/2}\|\boldsymbol\Omega\|_{\infty,\infty}^{2(1-\omega)}a_3''^{s-2}d_1''^{2^{s-2}}s_0(\log p)^{1-\omega/2}v(p)N_s^{\omega/2-1/2}=o(1),\\
\end{align*}
and for any $1\le s\le m$,
\begin{align*}
&\|\boldsymbol\Omega\|_{\infty,\infty}a_3''^{2s-2}d_1''^{2^{s}}sN_s^{\alpha_1-1/2}=o(1).
\end{align*}

\end{itemize}
Conditions (E9)  and (E10) are similar to
conditions (E7)  and (E8). In condition (E10), $g(s,s_0)=s^2\tilde M^{2^{s+1}}s_0$.
The following corollary demonstrates the asymptotic properties
of $\sqrt{N_s}(\hat\beta^{da(s)}_{l}-\beta_{l}^*)$.
\begin{corollary}
\label{C4}
Under the  conditions of Corollary \ref{C3}, suppose
that conditions (D1), (E9)  and (E10) are satisfied and for any $1\le s\le m$,
$h_s=c_{3s}''s\tilde M^{2^s}s_0^{1/2}\|\boldsymbol\Omega\|_{\infty,\infty}\sqrt{\log p/N_s}$  and $\kappa_s=c_{4s}''s\tilde M^{2^s}s_0^{1/2}\|\boldsymbol\Omega\|_{\infty,\infty}\sqrt{\log p/N_s}$, where $c_{3s}''$  and $c_{4s}''$ could be any constants no less than $1$. Then for any $1\le s\le m$ and   $1\le l\le p$, we have $\sigma_{ l}^{-1}\sqrt{N_s}(\hat\beta^{da(s)}_{l}-\beta_{l}^*)\longrightarrow N(0,1)$ in distribution as $p\to \infty$.
\end{corollary}

\section{Simulation studies}

In this section, we conduct extensive simulation studies to examine the finite-sample performance of the proposed online lasso and debiasing procedures.

\subsection{Evaluation of the online consistent estimation}

In this subsection, we first investigate the performance of the proposed online lasso method and randomly generate a total $N_m$ samples arriving in a sequence of $m$ data batches, denoted by $\left\{\mathcal{D}_{1}, \ldots, \mathcal{D}_{m}\right\}$, from the following two 
examples with the continuous and discrete outcome described in Section 3:\\
{\bf Model 1}: $Y_i^{(j)}=3 \boldsymbol X_i^{(j)\top}\boldsymbol\beta_{0}+10 \sin (\boldsymbol X_i^{(j)\top}\boldsymbol\beta_{0} )+\epsilon_i^{(j)}, i=1, \ldots, n_{j},~  j=1, \ldots, m,$\\
where $\boldsymbol{X}_{i}^{(j)}$ is generated from a multivariate normal distribution $\mathcal{N}(\boldsymbol{0},\boldsymbol{\Sigma})$
with covariance matrix $\boldsymbol{\Sigma}=\left(2^{-|k_1-k_2|}\right)_{1\leq k_1,k_2\leq p}$, and the true parameter 
$\boldsymbol{\beta}_0=\boldsymbol{\tilde\beta}/\|\boldsymbol{\Sigma}^{1/2}\boldsymbol{\tilde\beta}\|_2$ with
$$
\tilde{\beta}_{l}=\left\{\begin{array}{l}
l, \quad \text { for } 1 \leq l \leq s_0, \\
0, \quad \text { for } s_0+1 \leq l \leq p.
\end{array}\right.
$$
The random error $\epsilon_i^{(j)}$ is generated from four types of distributions: (i) standard normal distribution, denoted as $\mathcal{N}(0,1)$; (ii) log-normal distribution with the log location parameter 0 and log shape parameter 1, denoted as $\mathrm{LN}(0,1)$; (iii) Student's $t$-distribution with 3 degrees of freedom, denoted as $t(3)$; (iv) Weibull distribution with shape parameter $0.5$ and scale parameter $0.5$, denoted as Weibull(0.5; 0.5).

\noindent{\bf Model 2}: ${\rm Pr}(Y_i^{(j)} \mid \boldsymbol{X}_i^{(j)})=\frac{\exp\{ \boldsymbol{X}_i^{(j)\top}\boldsymbol{\beta}_{0}+\sin ( \boldsymbol{X}_i^{(j)\top})\boldsymbol{\beta}_{0}\}}{1+\exp \{ \boldsymbol{X}_i^{(j)\top}\boldsymbol{\beta}_{0}+\sin ( \boldsymbol{X}_i^{(j)\top}\boldsymbol{\beta}_{0})\}}, i=1, \ldots, n_{j},~ j=1, \ldots, m,$\\
where $\boldsymbol{X}_{i}^{(j)}$ is generated from a multivariate normal distribution $\mathcal{N}(\boldsymbol{0},\boldsymbol{\Sigma})$
with the same true parameter $\boldsymbol{\beta}_0$ as Model 1.
For the
design matrix, we consider two scenarios: (i) $\boldsymbol{\Sigma}$ is Toeplitz with ${\Sigma}_{k_1,k_2}=0.5^{|k_1-k_2|}$; (ii) $\boldsymbol{\Sigma} = \boldsymbol{I}$.
For each type of model, we consider the following combinations of $(N_m,m,n_j,p,s_0)$, $j=1,\ldots,m$: (i) $(N_m,m,n_j,p,s_0) = (1600, 16, 100, 200, 5)$; (ii) $(N_m,m,n_j,p,s_0) = (3200, 16,$ $ 200,400, 10)$.

For comparison, we consider the following methods: (i) the proposed online lasso estimator at several intermediate points for $s=1,\ldots,m$, denoted by ``online''; (ii) the offline lasso estimator at the terminal time point $m$, denoted by ``offline''; (iii) the offline lasso estimator with final data batch $\mathcal{D}_m$,
denoted by ``final''.  To measure the estimation accuracy, we calculate the sine distance between the estimator $\hat{\beta}_{\tau}$ and true parameter $\beta_{0}$ defined as follows:
$$
\sin \theta\left(\hat{\beta}_{\tau}, \beta_{0}\right)=1-\frac{<\hat{\beta}_{\tau}, \beta_{0}>}
{\|\hat{\beta}_{\tau}\|_{2}\left\|\beta_{0}\right\|_{2}},
$$
where $<a, b>$ is the inner product of vectors $a$ and $b$. Here we report the sine distance rather than $\|\hat{\beta}_{\tau}-c_{\tau} \beta_{0}\|_{2}$ for all simulation configurations.
As $c_{\tau}$ may take different values under different model and different settings, the sine distance is free of $c_{\tau}$.

The tuning parameters $\lambda_s$ and $\gamma_s$, $s=1,\ldots,m$, are chosen by the modified BIC \citep{wang2007tuning}.
For example, we obtain $\lambda_s$ by minimizing
\begin{align*}
\operatorname{BIC}(\lambda_s)=&
\log \bigg[ (\hat{\boldsymbol\beta}(\lambda_s)-\hat{\boldsymbol\beta}_2^{(s-1)})^\top
\sum_{j=1}^{s-1}\frac{n_j }{2N_s}  \boldsymbol H_1^{(j)}(\hat{\boldsymbol\beta}(\lambda_s)-\hat{\boldsymbol\beta}_2^{(s-1)})
\\& + \frac{2 }{N_s}  \sum_{i=1}^{n_s/2}l(Y_i^{(s)},\boldsymbol X_i^{(s)\top}\hat{\boldsymbol\beta}(\lambda_s)) \bigg]
 +C_{N_s} \frac{\log (N_s/2)}{N_s/2}
\big\|\hat{\boldsymbol\beta}(\lambda_s)\big\|_{0},
\end{align*}
where $\hat{\boldsymbol\beta}(\lambda_s)$ is obtained from (\ref{E4}),
$C_{N_s}=c \log \log (p)$, $c$ is a constant, and $\|\cdot\|_{0}$ denotes the number of nonzero elements in a vector.
Furthermore, we choose the robustification parameter $\tau$ in Huber loss such that $80 \%$ of the prediction errors are in $[-\tau, \tau]$.

Tables \ref{Simulation-Results-consistency} summarizes the results for Models 1 and 2 averaged over $200$ replications. It can be see the sine distance of the proposed online estimator decreases rapidly as the number of data batches $s$ increasing from 1 to 16 across different settings. Meanwhile, the performance of the proposed online estimator is close to the offline benchmark method, which implies the feasibility of the proposed online procedure. In addition, one can see that the proposed online method is robust to different types of error terms for the Huber loss function. Overall, the proposed online method performs satisfactorily with a relatively small sine distance in all settings.

\begin{table}[htb]
\caption{The sine distance under different settings in Section 4.1 are summarized over
200 replications.
} \label{Simulation-Results-consistency}
\vspace{0.25cm}
\renewcommand\arraystretch{1.25}
\centering
\begin{tabular}{  c cccc  cccc}
\hline
  &   & \multicolumn{4}{c}{ online}  & offline    & final \\
Model&Batch index $s$ & 4& 8 &  12  & 16 \\
\hline
   \multicolumn{8}{c}{ $(N_{m}, m, n_j, p, s_0)=(1600, 16, 100, 200, 5)$}
   \\ \hline
 &$\mathcal{N}$(0,1) & 0.055 &	0.030 &	0.013 &	0.007 &	0.007 &	0.071

 \\
Model 1 &LN(0,1)& 0.102 & 	0.064 & 	0.026 & 	0.013&  	0.012 	& 0.107

 \\
 &$t$(3)  &0.072 &	0.040& 	0.017 &	0.008 	&0.009& 	0.090

 \\
 &Weibull(0.5,0.5)  &0.089 	&0.063 &	0.028 &	0.014 	&0.018 &	0.117
\\ \hline
   \multicolumn{8}{c}{ $(N_{m}, m, n_j, p, s_0)=(3200, 16, 200, 400, 10)$}
 \\ \hline
 &$\mathcal{N}$(0,1) & 0.089 &	0.049 &	0.020 &	0.009 &	0.011 &	0.138

 \\
Model 1 &LN(0,1)& 0.144 &	0.090 &	0.037 &	0.018 &	0.019 &	0.195

 \\
 &$t$(3)  & 0.116 &	0.070 &	0.028 &	0.013 &	0.015 &	0.171

 \\
 &Weibull(0.5,0.5)  &0.159 &	0.102& 	0.042 &	0.020 &	0.022 &	0.215

 \\
\hline
   \multicolumn{8}{c}{ $(N_{m}, m, n_j, p, s_0)=(1600, 16, 100, 200, 5)$}
 \\ \hline

Model 2 &$\Sigma=I$ & 0.183& 	0.083 &	0.060 & 	0.052& 	0.038 &	0.371

 \\
 &$\Sigma=(0.5^{|k_1-k_2|})$ &0.113 &	0.064 &	0.052 &	0.049 &	0.038 &	0.340

 \\ \hline
   \multicolumn{8}{c}{ $(N_{m}, m, n_j, p, s_0)=(3200, 16, 200, 400, 10)$}
 \\ \hline

 Model 2 &$\Sigma=I$   &0.165 &	0.078 &	0.057 &	0.049 &	0.035 &	0.339

 \\
 &$\Sigma=(0.5^{|k_1-k_2|})$  &0.117 &	0.070 &	0.055 &	0.048 	&0.040 	&0.339

 \\
\hline

\end{tabular}

\end{table}

\subsection{Evaluation of the online pointwise inference}

In this subsection, we conduct simulations to check the performance of the online debiasing estimator via the null hypothesis $H_{0, l}: \beta_{l}^{*}=0$, $l\in\{1,\ldots,p\}$, which is equivalent to the null hypothesis $H_{0, l}: \beta_{0, l}=0$. We consider two types of examples under the same settings as in the first part except for the different combinations of $(N_m,m,n_j,p,s_0)$, $j=1,\ldots,m$: (i) $(N_m,m,n_j,p,s_0) = (1600, 16, 100, 200, 5)$; (ii) $(N_m,m,n_j,p,s_0) = (2400, 12, 200,$ $400, 10)$.

For comparison, we consider the following methods: (i) the proposed online debiasing estimator at several intermediate points for $s=1,\ldots,m$, denoted by ``online-deb''; (ii) the offline debiasing estimator at the terminal time point $m$, denoted by ``offline-deb''; (iii) the offline debiasing estimator with final data batch $\mathcal{D}_m$,
denoted by ``final-deb''. To evaluate the performance of different methods, we compute the following measurements:

\noindent (a) FPR: the average False Positive Rate corresponding to zero coefficients $\beta_{l}, s_0+1 \leq l \leq p$;

\noindent (b) $\operatorname{TPR}({l})$: the True Positive Rate corresponding to
$\beta_{l}, 1 \leq l \leq s_0$.

The detailed calculations for  the $s$th batch are given by
\begin{align*}
\text{FPR} = & \text{Average} \Big\{
\frac{1}{p-s_0} \sum_{l=s_0+1}^p I \big( \sqrt{N_s} |\hat\beta^{da(s)}_{l}|/ \hat\sigma_{l,s}  \geq z_{\alpha/2} \big)  \Big\},
\\
\operatorname{TPR}({l}) = & \text{Average}  \Big\{
 I\big(  \sqrt{N_s} |\hat\beta^{da(s)}_{l}|/ \hat\sigma_{l,s}  \geq z_{\alpha/2} \big)\Big\},
\end{align*}
where 
``Average" represents the average rate  over $200$ replications.

The tuning parameters $h_s$ and $\kappa_s$, $s = 1, \cdots , m,$ are determined as follows. Following \cite{cai2011constrained}, we can use the offline cross-validation scheme to select the tuning parameters $h_1$ and
$\kappa_1$ in (\ref{inver}) with only the first data batch $\mathcal{D}_1$. However, it is infeasible for streaming data since we can not access the entire raw data at the same time. Motivated by \cite{tashman2000out} and \cite{han2021online}, we adopt the following ``rolling-original-recalibration'' scheme to select the tuning parameters $h_s$, $\kappa_s$, $s=1,\ldots,m$. Here, we just present the selection of $h_s$, the similar idea can be used for $\kappa_s$. For $s\geq 2$, we regard the previous cumulative data set $\{\mathcal{D}_1,\ldots,\mathcal{D}_{s-1}\}$ as the training set that trains the estimator $\widehat{\bm\Omega}_1^{(s-1)}(h)$ for a sequence of $h$ in a candidate set $\mathcal{S}_h$ while the current data batch $\mathcal{D}_s$ is the validation set. Thus, when the data batch $\mathcal{D}_s$ arrives, we select $h_s$ by choosing the smallest likelihood loss on the validation sample as follows:
\begin{align*}
    h_s = \mathop{\arg\min}_{h\in\mathcal{S}_h}\left({\rm{tr}}\left\{2{\boldsymbol{H}}_1^{(s)}\widehat{\bm\Omega}_1^{(s-1)}(h)/n_s\right\} - \log[\det\{\widehat{\bm\Omega}_1^{(s-1)}(h)\}]\right).
\end{align*}

Tables \ref{Simulation-Results-pointwise-inference-huber1}--\ref{Simulation-Results-pointwise-inference-logistics2}
summarize the FPR and TPRs for the proposed online pointwise tests at the significance level of $0.05$ over 200 replications. We find that
all the methods preserve nominal levels across different settings. In the meanwhile, the empirical powers (TPRs) of the proposed online method perform better than the final-deb method, which implies the advantage of the proposed online method.

\begin{table}[htb]
\caption{The average True/False positive rates under different settings for Model 1 with $(N_m,m,n_j,p,s_0) = (1600, 16, 100, 200, 5)$ in Section 4.2 are summarized over
200 replications.
} \label{Simulation-Results-pointwise-inference-huber1}
\vspace{0.25cm}
\renewcommand\arraystretch{1.25}
\centering
\begin{tabular}{  c cccc  cccc}
\hline
  &   & \multicolumn{4}{c}{online-deb}  & offline-deb    & final-deb \\
&Batch index $s$ & 4& 8 &  12 & 16  \\
\hline

 &FPR & 0.045 &	0.044 &	0.050 &	0.050 &	0.053 & 0.050

 \\
 &TPR(1)& 1.000 & 	1.000 & 	1.000& 	1.000&  	1.000 & 1.000

 \\
 &TPR(2)  &1.000 &1.000& 	1.000 &1.000 	&1.000& 1.000

 \\
$\mathcal{N}$(0,1)&TPR(3)  &1.000 	&1.000 &	1.000 &	1.000 	&1.000 & 1.000
\\
 &TPR(4) & 1.000&1.000 &1.000 &	1.000 &1.000 &1.000
\\
 &TPR(5) & 0.965 &	1.000 &	1.000 &	1.000 &1.000 & 0.910

 \\ \hline
 &FPR & 0.046 &	0.045 &	0.050 &	0.053 &	0.053 & 0.052

 \\
 &TPR(1)& 1.000 & 	1.000 & 	1.000 & 	1.000&  	1.000 	& 1.000

 \\
 &TPR(2)  &1.000 &	1.000& 	1.000 &	1.000 	&1.000& 1.000

 \\
LN(0,1)&TPR(3)  &1.000	&1.000 &	1.000 &1.000 & 1.000	&1.000
\\
 &TPR(4) & 1.000 &	1.000 &	1.000 &	1.000 &	1.000 & 1.000
\\
 &TPR(5) & 0.955 &	1.000 &	1.000 &	1.000 &	1.000 & 0.880

 \\ \hline

\end{tabular}

\end{table}

\begin{table}[htb]
\caption{The average True/False positive rates under different settings for Model 2 with $(N_m,m,n_j,p,s_0) = (1600, 16, 100, 200, 5)$ in Section 4.2 are summarized over
200 replications.
} \label{Simulation-Results-pointwise-inference-huber2}
\vspace{0.25cm}
\renewcommand\arraystretch{1.25}
\centering
\begin{tabular}{  c cccc  cccc}
\hline
 &   & \multicolumn{4}{c}{online-deb}  & offline-deb    & final-deb \\
 $\boldsymbol{\Sigma}$&Batch index $s$ & 4& 8 &  12 & 16  \\
\hline

 &FPR & 0.038 &	0.047 &	0.050 &	0.048 &	0.048 & 0.043

 \\
 &TPR(1)& 1.000 & 	1.000 & 	1.000& 	1.000&  	1.000 & 0.990

 \\
 &TPR(2)  &1.000 &1.000& 	1.000 &1.000 	&1.000& 0.845

 \\
$\boldsymbol{I}$&TPR(3)  &0.980 	&1.000 &	1.000 &	1.000 	&1.000 & 0.595
\\
 &TPR(4) & 0.720 &0.970 &1.000 &	1.000 &1.000 &0.315
\\
 &TPR(5) & 0.225 &	0.555 &	0.760 &	0.850 &0.930 & 0.105

 \\ \hline
 &FPR & 0.044 &	0.047 &	0.049 &	0.052 &	0.048 & 0.045

 \\
 &TPR(1)& 1.000 & 	1.000 & 	1.000 & 	1.000&  	1.000 	& 0.990

 \\
 &TPR(2)  &1.000 &	1.000& 	1.000 &	1.000 	&1.000& 0.995

 \\
$(0.5^{|k_1-k_2|})$&TPR(3)  &0.955	&1.000 &	1.000 &1.000 & 1.000	&0.985
\\
 &TPR(4) & 0.670 &	0.910 &	0.985 &	1.000 &	0.995 & 0.700
\\
 &TPR(5) & 0.250 &	0.510 &	0.685 &	0.780 &	0.635 & 0.315

 \\ \hline

\end{tabular}

\end{table}

\begin{table}[htb]
\caption{The average True/False positive rates under different settings for Model 1 with $(N_m,m,n_j,p,s_0) = (2400, 12, 200,$ $400, 10)$ in Section 4.2 are summarized over
200 replications.
} \label{Simulation-Results-pointwise-inference-logistics1}
\vspace{0.25cm}
\renewcommand\arraystretch{0.95}
\centering
\begin{tabular}{  c cccc  cccc}
\hline
  &   & \multicolumn{4}{c}{online-deb}  & offline-deb    & final-deb \\
&Batch index $s$ & 3& 6 &  9 & 12  \\
\hline

 &FPR & 0.046 &	0.046 &	0.049 &	0.050 &	0.050 & 0.049

 \\
 &TPR(1)& 1.000 & 	1.000 & 	1.000& 	1.000&  	1.000 & 1.000

 \\
 &TPR(2)  &1.000 &1.000& 	1.000 &1.000 	&1.000& 1.000

 \\
&TPR(3)  &1.000 	&1.000 &	1.000 &	1.000 	&1.000 & 1.000
\\
 &TPR(4) & 1.000&1.000 &1.000 &	1.000 &1.000 &1.000
\\
$\mathcal{N}$(0,1) &TPR(5) & 1.000 &	1.000 &	1.000 &	1.000 &1.000 & 1.000
\\
 &TPR(6)& 1.000 &	1.000 &	1.000 &	1.000 &1.000 & 1.000
 \\
 &TPR(7) & 1.000 &	1.000 &	1.000 &	1.000 &1.000 & 0.990
 \\
 &TPR(8) & 0.975 &	1.000 &	1.000 &	1.000 &1.000 & 0.920
 \\
 &TPR(9) & 0.760 &	0.925 &	0.965 &	1.000 &1.000 & 0.680
 \\
 &TPR(10) & 0.350 &	0.480 &	0.680 &	0.800 &0.800 & 0.370
 \\ \hline
 &FPR & 0.046  &	0.047 &	0.050 &	0.050 &	0.050 & 0.049

 \\
 &TPR(1)& 1.000 & 	1.000 & 	1.000 & 	1.000&  	1.000 	& 1.000

 \\
 &TPR(2)  &1.000 &	1.000& 	1.000 &	1.000 	&1.000& 1.000

 \\
&TPR(3)  &1.000	&1.000 &	1.000 &1.000 & 1.000	&1.000
\\
 &TPR(4) & 1.000 &	1.000 &	1.000 &	1.000 &	1.000 & 1.000
\\
LN(0,1) &TPR(5) & 1.000 &	1.000 &	1.000 &	1.000 &	1.000 & 1.000
\\
 &TPR(6) & 1.000 &	1.000 &	1.000 &	1.000 &	1.000 & 1.000
 \\
 &TPR(7) & 1.000 &	1.000 &	1.000 &	1.000 &	1.000 & 0.995
 \\
 &TPR(8) & 0.965 &	1.000 &	1.000 &	1.000 &	1.000 & 0.900
 \\
 &TPR(9) & 0.700 &	0.880 &	0.965 &	0.995 &	1.000 & 0.640
 \\
 &TPR(10) & 0.305 &	0.480  &	0.625 & 0.745 &	0.750 & 0.330
 \\ \hline

\end{tabular}

\end{table}

\begin{table}[htb]
\caption{The average True/False positive rates under different settings for Model 2 with $(N_m,m,n_j,p,s_0) = (2400, 12, 200,$ $400, 10)$ in Section 4.2 are summarized over
200 replications.
}\label{Simulation-Results-pointwise-inference-logistics2}
\vspace{0.25cm}
\renewcommand\arraystretch{0.95}
\centering
\begin{tabular}{  c cccc  cccc}
\hline
  &   & \multicolumn{4}{c}{online-deb}  & offline-deb    & final-deb \\
$\boldsymbol{\Sigma}$&Batch index $s$ & 3& 6 &  9 & 12  \\
\hline

 &FPR & 0.041 &	0.049 &	0.050 &	0.050 &	0.050 & 0.043

 \\
 &TPR(1)& 1.000 & 	1.000 & 	1.000& 	1.000&  	1.000 & 0.990

 \\
 &TPR(2)  &1.000 &1.000& 	1.000 &1.000 	&1.000& 0.975

 \\
&TPR(3)  &1.000 	&1.000 &	1.000 &	1.000 	&1.000 & 0.900
\\
 &TPR(4) & 0.985 &1.000 &1.000 &	1.000 &1.000 &0.785
\\
$\boldsymbol{I}$ &TPR(5) & 0.970 &	1.000 &	1.000 &	1.000 &1.000 & 0.745
\\
 &TPR(6)& 0.850 &	0.990 &	1.000 &	1.000 &1.000 & 0.485
 \\
 &TPR(7) & 0.625 &	0.975 &	1.000 &	1.000 &1.000 & 0.385
 \\
 &TPR(8)  & 0.390 &	0.875 &	1.000 &	0.990 &1.000 & 0.190
 \\
 &TPR(9) & 0.250 &	0.510 &	0.670 &	0.710 & 0.830 &  0.160
 \\
 &TPR(10) & 0.085 &	0.016 &	0.200 &	0.290 & 0.375 &  0.080
 \\ \hline
 &FPR & 0.046 &	0.047 &	0.048 &	0.046 &	0.049 & 0.046

 \\
 &TPR(1)& 0.995 & 	1.000 & 	1.000 & 	1.000&  	1.000 	& 0.935

 \\
 &TPR(2)  &0.995 &	1.000& 	1.000 &	1.000 	&1.000& 0.980

 \\
&TPR(3)  &0.960	& 0.995 &	1.000 &1.000 & 1.000	&0.980
\\
 &TPR(4) & 0.945 &	0.995 &	1.000 &	1.000 &	1.000 & 0.940
\\
$(0.5^{|k_1 - k_2|})$ &TPR(5) & 0.875 &	1.000 &	1.000 &	1.000 &	1.000 & 0.865
\\
 &TPR(6) &0.715 &	0.945 &	1.000 &	1.000 &	1.000 & 0.775
 \\
 &TPR(7) &0.570 &	0.935 &	0.995 &	1.000 &	1.000 & 0.605
 \\
 &TPR(8) &0.395 &	0.640 &	0.820 &	0.930 &	0.925 & 0.350
 \\
 &TPR(9) &0.185 &	0.345 &	0.490 &	0.650 &	0.675 & 0.180
 \\
 &TPR(10) & 0.090 &	 0.225 & 0.305	 & 0.370  & 0.305  & 0.125
 \\ \hline

\end{tabular}

\end{table}

\section{Real data example}
\subsection{Nasdaq stock data}

In this subsection, we illustrate the proposed method with the Nasdaq stock dataset,
which is collected from January 1, 2008 to November 2, 2018.
For this dataset, the response variable is the return of the Nasdaq 100 index for every three days, and the covariates are $p=226$ stocks returns for every three days during this period.
Similar to \citet{lan2016testing}, our goal in this study is to find the most relevant stocks that can be used to construct a small portfolio, which tracks the return of the Nasdaq 100 index.

\begin{figure}[htbp]
\centering
\includegraphics[height = 10.1cm, width=14cm]{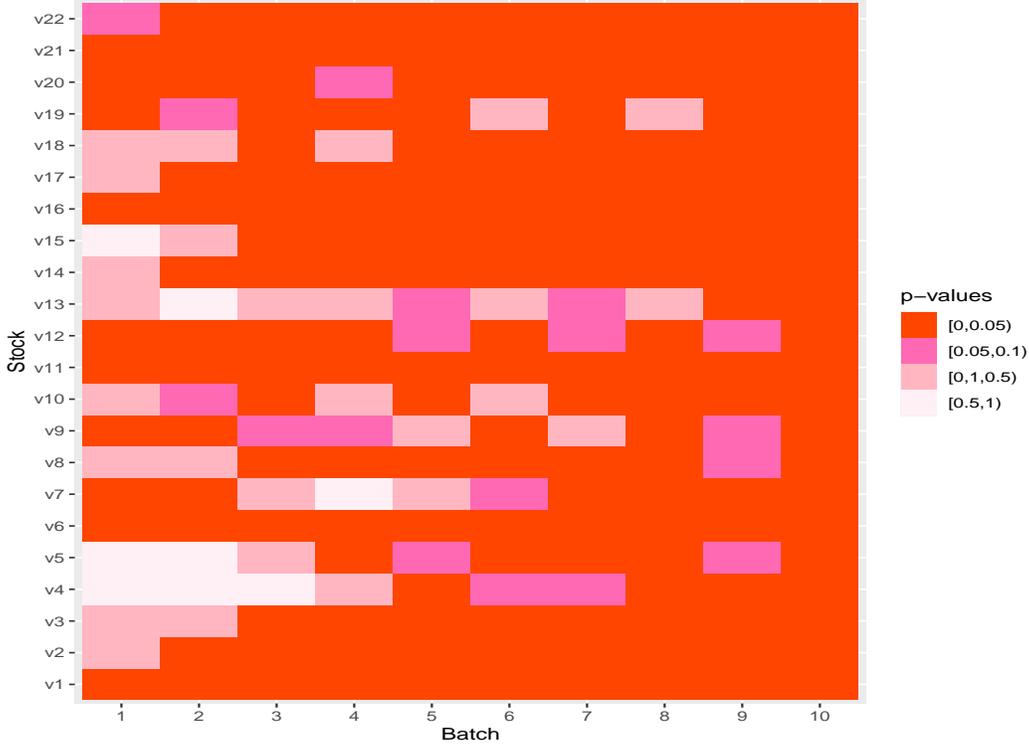}
\caption{results of $p$-values for Nasdaq stock data}
\label{real_data_Nasdaq_stock_p_values}
\end{figure}

To apply our proposed procedure,  the data are splitted into $m=10$ batches. We take the first two-year dataset as the first data batch ($n_1=164$) to guarantee a sufficiently large sample size at the initial stage and the next one year dataset as the subsequent data batch ($n_j=82, j=2,\cdots,m-1$) . In addition, the sample size of the final batch is $n_m=72$.  Hence, the streaming data consists of $m = 10$ data batches with a total sample size $N_m = 892$.
To identify important stocks that are associated with Nasdaq 100 index, we apply the proposed online procedure to sequentially test the significance of each regression coefficient at a prespecified level $\alpha=0.05$, i.e., testing $H_{0,l}:$ $\beta_{0,l} = 0$ for $l = 1,\ldots,p$. The selection methods of the tuning parameters $\lambda_s$, $\gamma_s$, $h_s$, and $\kappa_s$, $s=1,\ldots,m$ are the same as those in the simulation studies.
To ensure the stability of selection  in this online framework,
the identified stocks are required to  be significant  at the level of $0.1$ for the $m-1$ batch.
It is reasonable that financial managers maybe track the stocks more time
and establish a portfolio cautiously, especially for the risk averse investors.
We find that 22 stocks are identified as important stocks at the significance level of $0.05$. Correspondingly, the $p$-values of these  regression coefficients over the $10$ batches are plotted in Figure \ref{real_data_Nasdaq_stock_p_values}.
From this figure,
as we collect data more and more,
the most selected stocks are more significant and  relatively stable.
This example demonstrates that our proposed method can be applied to analyze the stocks dataset and perform reasonable well.

\subsection{Financial distress data}

In this section, we illustrate our method with the financial distress dataset, which is available from \textsf{https://www.kaggle.com/datasets/shebrahimi/financial-distress}. This dataset is collected from a sample of companies. Time series varies between $1$ to $10$ for each company. For this dataset, the financial distress index can be regarded as the response variable and other $82$ variables are covariates that consist of some financial and non-financial characteristics of the sampled companies. In addition, this dataset consists of a total of $N_m=1008$ observations, and the response and the covariates have been standardized to have zero mean and unit variance. Our goal of this study is to select the variables that significantly affect the companies' financial distress.

\begin{figure}[htbp]
\centering
\includegraphics[width=15cm]{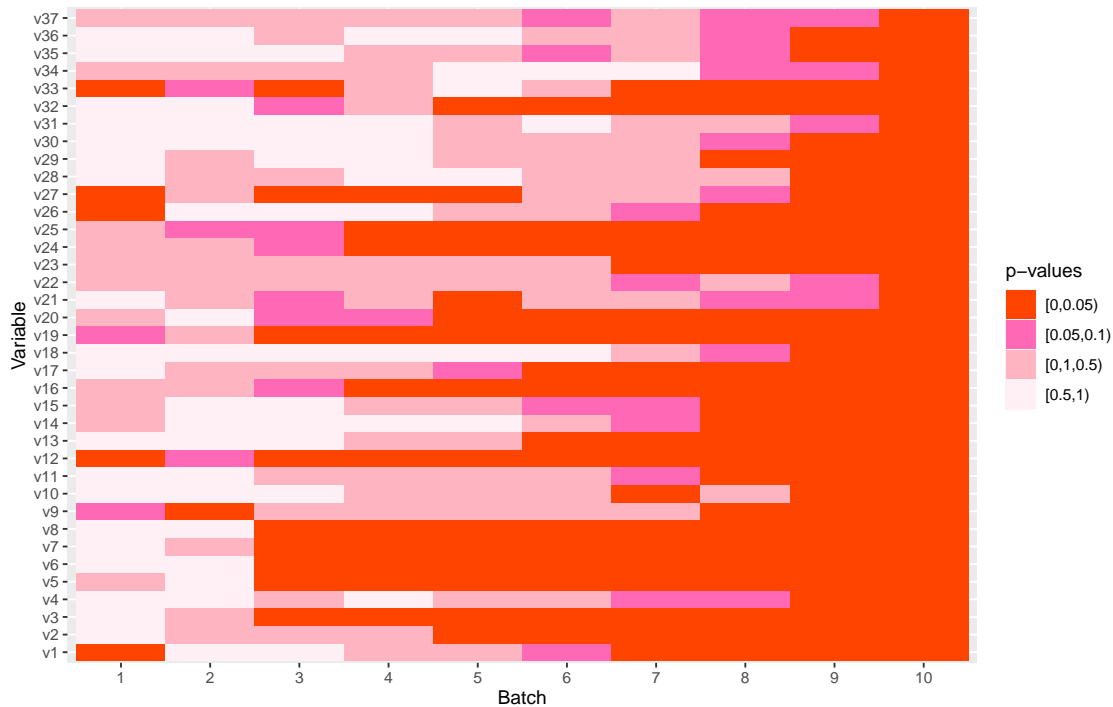}
\caption{results of $p$-values for financial distress data}
\label{real_data_financial_distress_p_values}
\end{figure}

In this example, the covariates include $190$ interaction terms (products of $20$ pairs of the original covariates). As a result, the dimension of the feature vector is $p = 272$.
Before applying our proposed procedure, we split the data into $m=10$ batches randomly, and take the $n_1=108$ observations as the first batch and set each of the  remaining $9$ batches contains $n_j=100$ observations. To identify the influential variables, we aim to test: $H_{0,l}:$ $\beta_{0,l} = 0$ for $l = 1,\ldots,p$. The tuning parameters $\lambda_s$, $\gamma_s$, $h_s$  and $\kappa_s$, $s=1,\ldots,m$ are determined by the same methods as described in the simulation studies. Given a prespecified level $\alpha=0.05$, we observe that 37 variables are significant in the online framework, and the associated $p$-values of the $10$ batches are presented in Figure \ref{real_data_financial_distress_p_values}.
From this figure, we can find that the most variables are
 more significant and
reach relative stability 
as more and more data are collected.
This example indicates 
that our proposed method can be applied to analyze the dataset with binary outcome and perform reasonable well.

\section{Discussion}

In this paper, we  studied  the statistical inference of SIMs with streaming data under the high-dimensional regime.
%
%
The proposed procedure was applicable to the streaming data, that is, only depended on the current batch of the data stream with summary statistics from the historical data. In addition, our method was developed for general convex loss functions, which could be effectively used to handle heavy-tailed errors or discrete responses. Meanwhile, we established the $\ell_1$ and $\ell_2$ bounds of the proposed online lasso estimators and the asymptotic normality of the proposed online debiased lasso estimators. Simulation
studies were conducted to show the effectiveness of the
proposed method and applications to two real data examples were
provided to illustrate our method.

There are several other interesting avenues for future
work. First, the current work relies on the assumption of the homogeneous data, that is, the streaming data is assumed to be i.i.d. sampled. It would be an interesting topic to address the problem of non-homogeneous data. Second, we require that the data is completely observed in our framework. It is unclear how to extend the proposed method in the presence of incomplete data such as missing data or censored data. Third, the selection of the parameter $\tau$ is crucial for the Huber loss function in real implementation. It is challenging to provide a data-driven selector to determine $\tau$ in a streaming manner with theoretical guarantees. We leave space here for future research.

\section*{Supplementary Material}

Supplementary material contains the proofs of main theorems with the required lemmas.

\newpage

\bibliographystyle{apalike}

\bibliography{reference}

\end{document}